\documentclass[twocolumn]{aastex631}

\usepackage{graphicx}
\usepackage{amsmath}
\usepackage[utf8x]{inputenc}
\usepackage[encapsulated]{CJK}
\usepackage{ucs}
\usepackage{chngcntr}
\graphicspath{{./}{figs/}}

\newcommand{\rd}{RUBIES-BLAGN-1}

\newcommand{\prospector}{\texttt{Prospector}}
\newcommand{\cloudy}{\texttt{Cloudy}}

\newcommand{\msun}{{\rm M}_{\odot}}

\newcommand{\lsun}{{\rm L}_\odot}

\newcommand{\zspec}{z_{\rm spec}}
\newcommand{\Av}{A_{\rm V}}

\newcommand{\he}{He\,\textsc{i}}
\newcommand{\ha}{H$\alpha$}
\newcommand{\pab}{Pa-$\beta$}
\newcommand{\pag}{Pa-$\gamma$}
\newcommand{\pad}{Pa-$\delta$}
\newcommand{\nii}{[N\,{\sc ii}]}
\newcommand{\oiii}{[O\,{\sc iii}]}
\newcommand{\cii}{[C\,{\sc ii}]}

\newcommand{\Ha}{H$\alpha$ }
\newcommand{\Pab}{Pa-$\beta$ }

\newcommand{\Nii}{[N\,{\sc ii}]}
\newcommand{\kms}{\rm km\,s^{-1}}

\shorttitle{Big Red Dot without Hot Dust}
\shortauthors{Wang et al}

\begin{document}

\title{RUBIES: JWST/NIRSpec Confirmation of an Infrared-luminous, Broad-line Little Red Dot with an Ionized Outflow}

\correspondingauthor{Bingjie Wang}
\email{bwang@psu.edu}

\author[0000-0001-9269-5046]{Bingjie Wang (\begin{CJK*}{UTF8}{gbsn}王冰洁\ignorespacesafterend\end{CJK*})}
\affiliation{Department of Astronomy \& Astrophysics, The Pennsylvania State University, University Park, PA 16802, USA}
\affiliation{Institute for Computational \& Data Sciences, The Pennsylvania State University, University Park, PA 16802, USA}
\affiliation{Institute for Gravitation and the Cosmos, The Pennsylvania State University, University Park, PA 16802, USA}

\author[0000-0002-2380-9801]{Anna de Graaff}
\affiliation{Max-Planck-Institut f\"ur Astronomie, K\"onigstuhl 17, D-69117, Heidelberg, Germany}

\author[0000-0002-3324-4824]{Rebecca L. Davies}
\affiliation{Centre for Astrophysics and Supercomputing, Swinburne University of Technology, Hawthorn, Victoria 3122, Australia}
\affiliation{ARC Centre of Excellence for All Sky Astrophysics in 3 Dimensions (ASTRO 3D), Australia}

\author[0000-0002-5612-3427]{Jenny E. Greene}
\affiliation{Department of Astrophysical Sciences, Princeton University, Princeton, NJ 08544, USA}

\author[0000-0001-6755-1315]{Joel Leja}
\affiliation{Department of Astronomy \& Astrophysics, The Pennsylvania State University, University Park, PA 16802, USA}
\affiliation{Institute for Computational \& Data Sciences, The Pennsylvania State University, University Park, PA 16802, USA}
\affiliation{Institute for Gravitation and the Cosmos, The Pennsylvania State University, University Park, PA 16802, USA}

\author[0000-0003-2680-005X]{Gabriel B. Brammer}
\affiliation{Cosmic Dawn Center (DAWN), Copenhagen, Denmark}
\affiliation{Niels Bohr Institute, University of Copenhagen, Jagtvej 128, Copenhagen, Denmark}

\author[0000-0003-4700-663X]{Andy D. Goulding}
\affiliation{Department of Astrophysical Sciences, Princeton University, Princeton, NJ 08544, USA}

\author[0000-0001-8367-6265]{Tim B. Miller}
\affiliation{Center for Interdisciplinary Exploration and Research in Astrophysics (CIERA) and Department of Physics \& Astronomy, Northwestern University, IL 60201, USA}

\author[0000-0002-1714-1905]{Katherine A. Suess}
\thanks{NHFP Hubble Fellow}
\affiliation{Kavli Institute for Particle Astrophysics and Cosmology and Department of Physics, Stanford University, Stanford, CA 94305, USA}

\author[0000-0001-8928-4465]{Andrea Weibel}
\affiliation{Department of Astronomy, University of Geneva, Chemin Pegasi 51, 1290 Versoix, Switzerland}

\author[0000-0003-2919-7495]{Christina C. Williams}
\affiliation{NSF's National Optical-Infrared Astronomy Research Laboratory, Tucson, AZ 85719, USA}
\affiliation{Steward Observatory, University of Arizona, Tucson, AZ 85721, USA}

\author[0000-0001-5063-8254]{Rachel Bezanson}
\affiliation{Department of Physics \& Astronomy and PITT PACC, University of Pittsburgh, Pittsburgh, PA 15260, USA}

\author[0000-0002-3952-8588]{Leindert A. Boogaard}
\affiliation{Max-Planck-Institut f\"ur Astronomie, K\"onigstuhl 17, D-69117, Heidelberg, Germany}

\author[0000-0001-7151-009X]{Nikko J. Cleri}
\affiliation{Department of Physics and Astronomy, Texas A\&M University, College Station, TX, 77843-4242 USA}
\affiliation{George P.\ and Cynthia Woods Mitchell Institute for Fundamental Physics and Astronomy, Texas A\&M University, College Station, TX, 77843-4242 USA}

\author[0000-0002-3301-3321]{Michaela Hirschmann}
\affiliation{Institute of Physics, Lab for galaxy evolution, EPFL, Observatoire de Sauverny, Chemin Pegasi 51, 1290 Versoix, Switzerland}

\author[0000-0003-1561-3814]{Harley Katz}
\affiliation{Sub-department of Astrophysics, University of Oxford, Oxford OX1 3RH, UK}

\author[0000-0002-2057-5376]{Ivo Labb\'e}
\affiliation{Centre for Astrophysics and Supercomputing, Swinburne University of Technology, Melbourne, VIC 3122, Australia}

\author[0000-0003-0695-4414]{Michael V. Maseda}
\affiliation{Department of Astronomy, University of Wisconsin-Madison, Madison, WI 53706, USA}

\author[0000-0003-2871-127X]{Jorryt Matthee}
\affiliation{Institute of Science and Technology Austria (ISTA), Am Campus 1, 3400 Klosterneuburg, Austria}

\author[0000-0002-2446-8770]{Ian McConachie}
\affiliation{Department of Physics and Astronomy, University of California, Riverside, Riverside, CA 92521, USA}

\author[0000-0003-3997-5705]{Rohan P. Naidu}
\thanks{NHFP Hubble Fellow}
\affiliation{MIT Kavli Institute for Astrophysics and Space Research, Cambridge, MA 02139, USA}

\author[0000-0001-5851-6649]{Pascal A. Oesch}
\affiliation{Department of Astronomy, University of Geneva, Chemin Pegasi 51, 1290 Versoix, Switzerland}
\affiliation{Cosmic Dawn Center (DAWN), Copenhagen, Denmark}

\author[0000-0003-4996-9069]{Hans-Walter Rix}
\affiliation{Max-Planck-Institut f\"ur Astronomie, K\"onigstuhl 17, D-69117, Heidelberg, Germany}

\author[0000-0003-4075-7393]{David J. Setton}
\thanks{Brinson Prize Fellow}
\affiliation{Department of Astrophysical Sciences, Princeton University, Princeton, NJ 08544, USA}

\author[0000-0001-7160-3632]{Katherine E. Whitaker}
\affiliation{Department of Astronomy, University of Massachusetts, Amherst, MA 01003, USA}
\affiliation{Cosmic Dawn Center (DAWN), Copenhagen, Denmark}

\begin{abstract}

The JWST discovery of ``little red dots'' (LRDs) is reshaping our picture of the early Universe, yet the physical mechanisms driving their compact size and UV-optical colors remain elusive. Here we report an unusually bright LRD ($\zspec=3.1$) observed as part of the RUBIES program. This LRD exhibits broad emission lines (FWHM$\sim4000\,\kms$), a blue UV continuum, a clear Balmer break and a red continuum sampled out to rest-frame 4~$\mu\mathrm{m}$ with MIRI. We develop a new joint galaxy and active galactic nucleus (AGN) model within the \texttt{Prospector} Bayesian inference framework and perform spectrophotometric modeling using NIRCam, MIRI, and NIRSpec/Prism observations. Our fiducial model reveals a $M_*\sim 10^9~\msun$ galaxy alongside a dust-reddened AGN driving the optical emission. Explaining the rest-frame optical color as a reddened AGN requires $\Av \gtrsim 3$, suggesting that a great majority of the accretion disk energy is re-radiated as dust emission. Yet, despite clear AGN signatures, we find a surprising lack of hot torus emission, which implies that either the dust emission in this object must be cold, or the red continuum must instead be driven by a massive, evolved stellar population of the host galaxy---seemingly inconsistent with the high EW broad lines (H$\alpha$ rest-frame EW $\sim 800$~\AA). The widths and luminosities of Pa-$\beta$, Pa-$\delta$, Pa-$\gamma$, and H$\alpha$ imply a modest black hole mass of $M_{\rm BH}\sim10^8$ M$_{\odot}$. Additionally, we identify a narrow blue-shifted \he\,$\lambda\,1.083\,\mu$m absorption feature in NIRSpec/G395M spectra, signaling an ionized outflow with kinetic energy up to $\sim 1$\% the luminosity of the AGN. The low redshift of RUBIES-BLAGN-1 combined with the depth and richness of the JWST imaging and spectroscopic observations provide a unique opportunity to build a physical model for these so-far mysterious LRDs, which may prove to be a crucial phase in the early formation of massive galaxies and their supermassive black holes
\footnote{Code implementing the joint galaxy and AGN model is made publicly available online as a part of \prospector; the model used in this work corresponds to the state of the Git repository
at commit \href{https://github.com/bd-j/prospector/pull/366}{1cce164}.}.

\end{abstract}


\section{Introduction}

Observations from the James Webb Space Telescope (JWST) have unveiled an intriguing sample of extremely red sources. As they typically also show a compact morphology these objects have been named ``little red dots" (LRDs; \citealt{Matthee2023:lrd}), and they are characterized by a red continuum in the rest-frame optical, a faint blue component in the rest-frame ultraviolet (UV), and often broad Balmer lines indicative of active galactic nuclei (AGNs). 
Spectroscopic follow-up of LRDs selected based on NIRCam colors has confirmed a surprising ubiquity of broad emission lines, strongly suggesting an AGN component in the rest-frame optical \citep{Greene2023,Labbe2023:agn}. 

At the same time, interpreting the spectral energy distributions (SEDs) of the LRDs remains difficult. It is still unclear how to model the different spectral components, and which parts of the SED are galaxy or AGN dominated. Common galaxy-only or AGN-only models struggle to simultaneously describe the blue slope in the rest-frame UV, and the very red slope toward the rest-frame optical. Early joint modeling of the two components (e.g., \citealt{Yang2020,Furtak2023,Williams2023,Vidal-Garcia2024}) has yielded ambiguous interpretations, while no consensus has been reached about whether the UV can be described by scattered AGN light \citep[e.g.,][]{Labbe2023:agn}, scattered galaxy light \citep[e.g.,][]{Killi2023}, or simply unobscured star formation \citep[e.g.,][]{Matthee2023:lrd}. 

The infrared (IR) colors pose additional puzzles: the red slope seen in LRDs strongly suggests high dust attenuation, which, when re-emitted in the IR, results in different spectral shapes depending on the AGN or galaxy nature of the source. AGN are typically associated with the presence of hot dust which will have a steeply rising red continuum at $>2\mu$m \citep{Lacy2004,Richards2006,Urrutia2012}, whereas the near-IR stellar bump at rest-frame 1.6~$\mu$m  leads to a bluer spectral slope \citep{Laurent2000,Sawicki2002}. A stacking analysis of LRD candidates in JADES \citep{Eisenstein2023} has shown a distinct lack of hot dust, yet no definitive conclusion is drawn regarding whether the flat MIRI color precludes the dominance of AGN in the rest optical, given the available data \citep{Williams2023}.
Broader samples selected by requiring both a red rest-optical and blue rest-UV continuum without a compactness criterion seem to be consistent with dusty star-formation \citep{Perez-Gonzalez2024}, sparking further debate between the galaxy or AGN origins of LRDs.
 
So far spectroscopic observations of LRDs have been limited to relatively small samples ($\lesssim$50; \citealt{Kocevski2023,Harikane2023,Matthee2023:lrd,Greene2023,Furtak2023b}), with the majority of available data restricted to JWST/NIRCam grism and/or JWST/Prism spectroscopy \citep{Greene2023,Kokorev2023,Killi2023}. The inferred number densities are quite high, comprising a few percent of the galaxy population at $z>5$ \citep[e.g.,][]{Matthee2023:lrd} and $\sim 20\%$ of the broad-line AGNs \citep[e.g.,][]{Harikane2023}. One lensed LRD presents a galaxy size limit of $<30$~pc, and a implied black hole to galaxy mass ratio of at least a few percent \citep{Furtak2023b}. Understanding how these massive black holes grew as early as $z=8.5$ \citep{Kokorev2023} in such compact galaxies remains an intriguing puzzle \citep[e.g.,][]{Pacucci2024,Silk2024}. 

The absence of medium to high-resolution spectra to date has meant that emission lines cannot be robustly decomposed, preventing robust measurements of gas kinematics. Accordingly, it is also difficult to characterize potential outflows, which are an integral part of AGN feedback \citep{Zakamska2014,Rupke2017,Davies2020,Laha2021}. As a consequence, it is unclear in which phase of supermassive black hole growth LRDs are. Although NIRCam/grism spectroscopy has provided emission line kinematics for a larger sample of LRDs \citep{Matthee2023:lrd}, these observations are limited in wavelength coverage, typically yielding one or two emission lines rather than a suite of diagnostics.

The elusive nature of LRDs has been further compounded by the recent discovery of co-existing Balmer breaks and broad Balmer emission lines in three LRDs at $z \sim 7-8$ \citep{Wang2024:bb}.
The Balmer breaks suggest that stellar emission dominates blueward of rest 0.4 $\mu$m, whereas the broad Balmer emission lines indicate emission from a dust-reddened AGN contributing to fluxes redward of rest 0.6 $\mu$m. 
Yet, the limiting cases of AGN/galaxy models explored in \citet{Wang2024:bb}, ranging from 0-100\% AGN contributions in the red continuum, fit the data equally well.

In this paper, we present a bright LRD at a spectroscopic redshift of $\zspec=3.1034 \pm 0.0002$ that is luminous in the IR and has multiple broad emission lines indicative of an AGN. \rd\ is discovered as part of the RUBIES program (JWST-GO-4233; PIs de Graaff \& Brammer; \citealt{deGraaff2024:survey}), for which we obtained high signal-to-noise spectra in both the JWST/NIRSpec low-resolution Prism and the medium-resolution G395M modes.
Notably, \rd\ is detected in JWST/MIRI F770W and F1800W imaging from PRIMER (JWST-GO-1837; PI Dunlop). The exquisite wavelength coverage allows for in-depth characterization of the physics of this object.

The structure of this paper is as follows. Section~\ref{sec:data} provides an overview of the data, including imaging, selection, and spectroscopy. Section~\ref{sec:data_analyses} details the SED modeling, emission-line decomposition, and photoionization modeling. Section~\ref{sec:res} presents the results on the inferred AGN and host galaxy properties, morphology, gas kinematics, and mass outflow rate. We conclude in Section~\ref{sec:concl} with discussion of the key findings.

Where applicable, we adopt the best-fit cosmological parameters from the WMAP 9 yr results: $H_{0}=69.32$ ${\rm km \,s^{-1} \,Mpc^{-1}}$, $\Omega_{M}=0.2865$, and $\Omega_{\Lambda}=0.7135$ \citep{Hinshaw2013}, and a \citet{Chabrier2003} initial mass function. Unless otherwise mentioned, we report the median of the posterior, and 1$\sigma$ error bars are the 16th and 84th percentiles.

\section{Data\label{sec:data}}

RUBIES is a spectroscopic survey with JWST/NIRSpec that targets $\sim 4000-5000$ NIRCam-selected sources across the UDS and EGS fields using the micro-shutter array (MSA). Observations are obtained with both the low-resolution Prism/CLEAR and medium-resolution G395M/F290LP modes, providing an unprecedented view of red sources at $z>3$. Unique to RUBIES is its highly complete sampling of the extremes in color-magnitude space.
\rd\ was selected with highest priority based on its extreme red color ($\rm F150W-F444W > 3.0~ mag$)---roughly 10 objects in CEERS \citep{Barrufet2023}, and merely 13 objects out of all observed in JADES-GOODS-S \citep{Williams2023} have such a red color.
It is subsequently targeted in two of the three masks observed in January 2024.
Full details of the survey design and data reduction is described in \citet{deGraaff2024:survey}. The specific observations analyzed here can be accessed via\dataset[DOI: 10.17909/c3t4-9p39]{gttps://doi.org/10.17909/c3t4-9p39}. This section provides a brief summary of the imaging and spectroscopic data.

\subsection{Imaging\label{subsec:img}}

\begin{figure*} 
\gridline{
  \fig{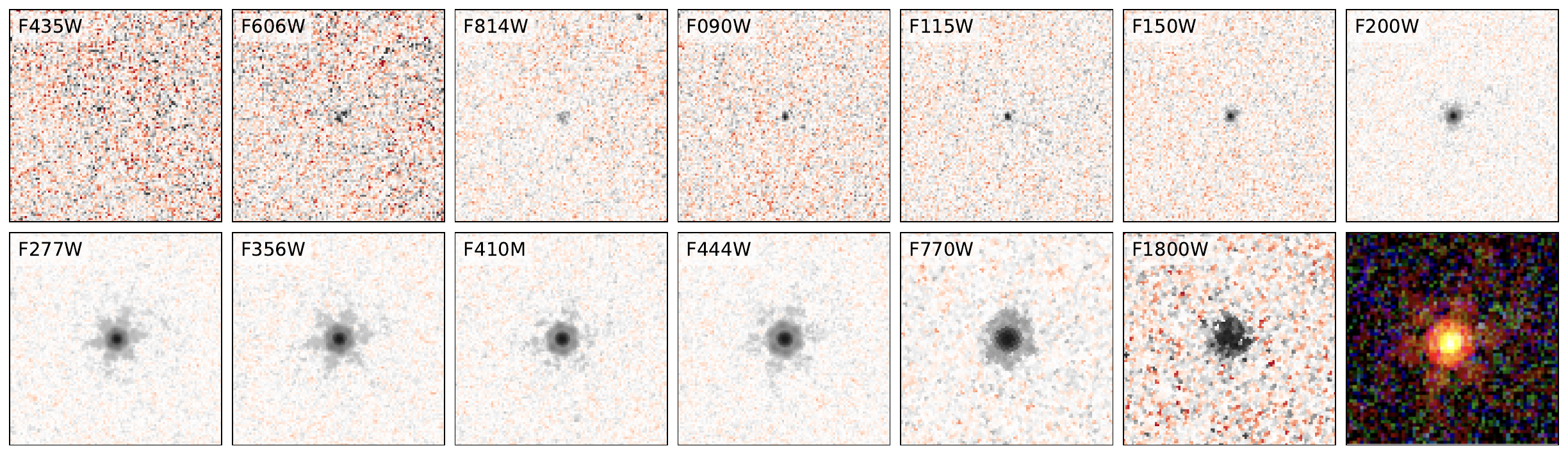}{0.95\textwidth}{(a)}
} 
 \gridline{
   \fig{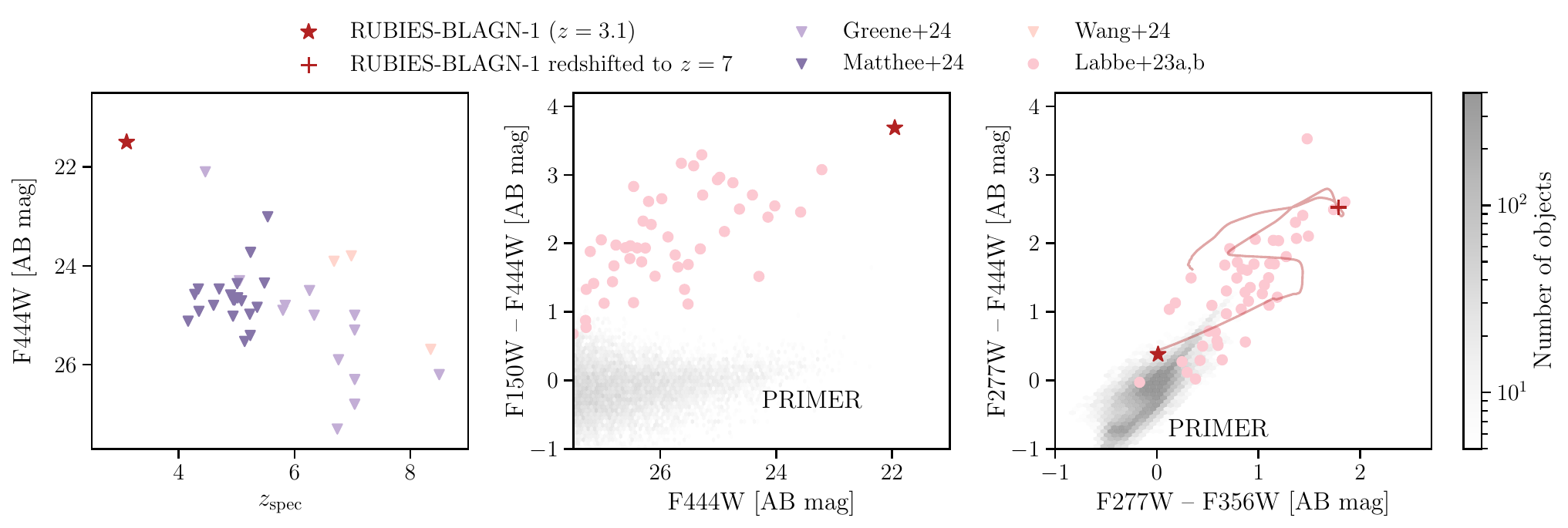}{0.95\textwidth}{(b)}
} 
\caption{Photometric observations of \rd. (a) Cutouts are shown in selected filter bands, including HST/WCS, JWST/NIRCam, and MIRI detections. A color image, composited from NIRCam/F115W, F277W, and F444W, is included as the last panel. Each panel is 5\arcsec\ in width. (b) \rd\ is spectroscopically confirmed to be among the lowest redshift LRDs. The spectroscopic redshifts of the samples from \citet{Greene2023,Matthee2023:lrd,Wang2024:bb} are plotted as light purple, purple, and pink triangles, respectively. \rd\ is bright in F444W ($\sim$ 22 AB mag) and exhibits an extreme red color with F150W-F444W $\sim$ 3.8 mag. The red objects at higher redshifts are included as pink dots for reference \citep{Labbe2023:massive,Labbe2023:agn}. The parent sample, including all sources identified from PRIMER having F444W $<$ 27.5 AB mag, is shown in gray. To facilitate a better comparison to the high-$z$ LRDs, we show a track tracing the colors of \rd\ as it moves to higher redshifts. Its color at $z=7$ is explicitly included as a red plus sign. \rd\ is clearly an outlier in the color space, and possibly can be seen as an analog of the high-$z$ LRDs.
}
\label{fig:color}
\end{figure*}

The RUBIES targets in the UDS were selected based on public JWST/NIRCam imaging from the PRIMER survey \citep{Donnan2024}, which provides NIRCam imaging in 8 bands (F090W, F115W, F150W, F200W, F277W, F356W, F444W and F410M) and MIRI imaging in the F770W and F1800W bands. 
Archival imaging from the Hubble Space Telescope (HST) is obtained from the CANDELS survey \citep{Grogin2011,Koekemoer2011}, which adds in 7 more bands in F435W, F606W, F814W, F105W, F125W, F140W, and F160W. 

We use the latest version (v7.0) of the publicly available image mosaics from the DAWN JWST Archive (DJA). These images were reduced using \texttt{grizli} \citep{grizli}, as also described in \citet{Valentino2023}, and have a pixel scale of $0.04\arcsec\,{\rm pix}^{-1}$. We run SourceExtractor \citep{Bertin1996} in dual image mode, using an inverse-variance weighted stack of F277W, F356W and F444W band as the detection image. Next, for each NIRCam band we construct a mosaic that is PSF-matched to the F444W image to measure fluxes in circular apertures with a radius of $0.16\arcsec$. We then scale the aperture fluxes to the flux measured through a Kron aperture on a PSF-matched version of the detection image, and to the total flux by dividing by the encircled energy of the Kron aperture on the F444W PSF \citep{Weibel2024}. 

Fluxes from the MIRI images are measured using a larger circular aperture of radius $0.35\arcsec$. 
The MIRI images were reduced in August 2023 (v7.0 on the DJA), prior to the release of new photometric calibrations. Following the JWST documentation, we therefore multiply the measured fluxes with a correction factor based on the updated reference file (\texttt{jwst\_miri\_photom\_0172}); a factor 0.85 and 1.03 for the F770W and F1800W bands, respectively. Finally, we use the empirical PSF models of \citet{Libralato2023} to scale the MIRI aperture fluxes to total fluxes. This is motivated by our morphological analysis (detailed in Appendix~\ref{app:morph}) which shows that \rd\ is unresolved in the long wavelength images, with a half-light size smaller that a single pixel, $R_{\rm e} < 0.25$~kpc.

Cutouts for all filters are presented in Figure~\ref{fig:color} together with the observed colors of \rd\ and the parent sample, highlighting the extreme color, magnitude, and compactness of the source. For reference, we also put \rd\ on a color-color plot of F277W-F356W vs. F277W-F444W---a criterion used in the literature to select red objects at $z \gtrsim 4$ \citep{Labbe2023:massive,Labbe2023:agn}. Clearly \rd\ is an outlier when compared to the parent PRIMER sample. Shifting the spectrum of \rd\ to $z=7$ reveals that its colors are similar to those of the high-$z$ LRDs, suggesting it may be taken as a lower-redshift analog.

In addition, we measure upper limits in Spitzer/MIPS 24 $\mu$m, Herschel/PACS 100, 160 $\mu$m, and Herschel/Spire 250, 350 $\mu$m, from the 3D-Herschel project (K. Whitaker et al. in prep). The imaging data is processed following \citet{Whitaker2014}, adopting a forced photometry methodology with the addition of both position and flux priors.

\subsection{Spectroscopy\label{subsec:spec}}

\begin{figure*} 
\gridline{
  \fig{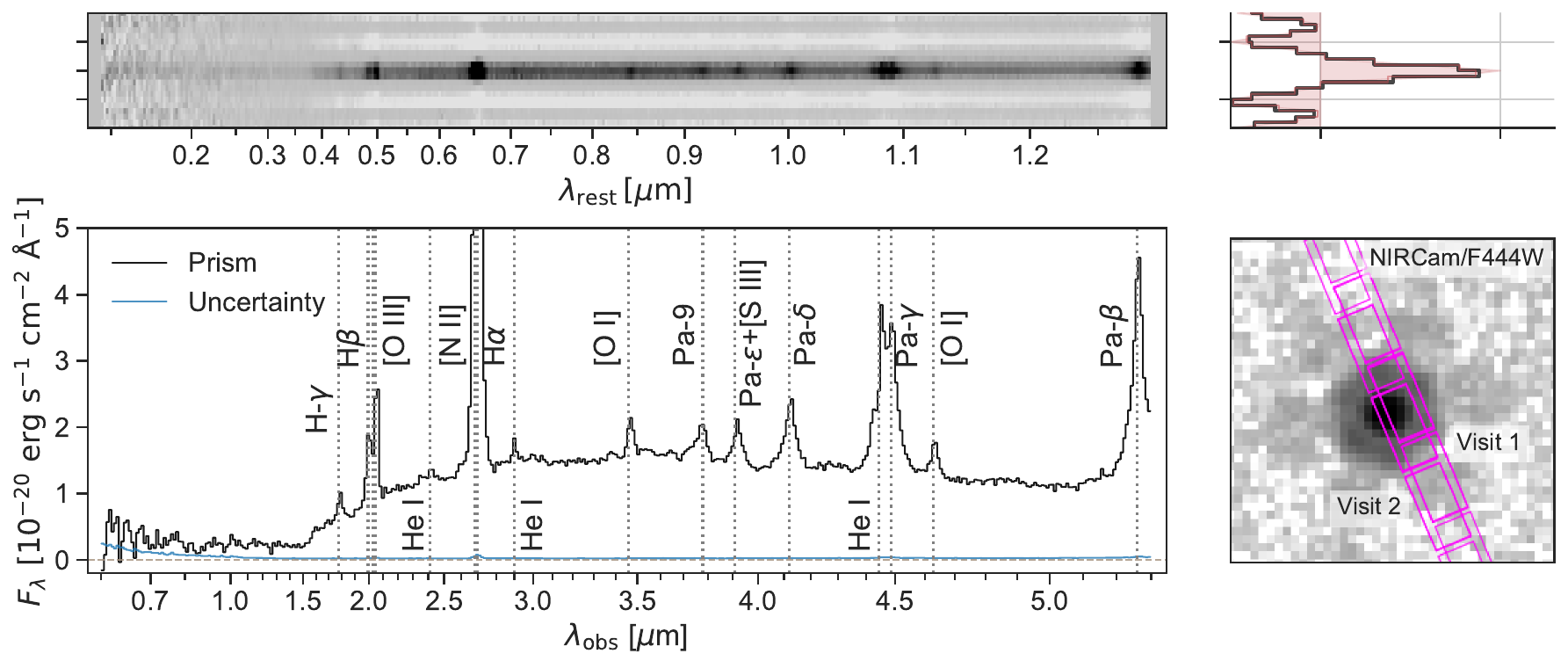}{0.95\textwidth}{(a)}
} 
 \gridline{
   \fig{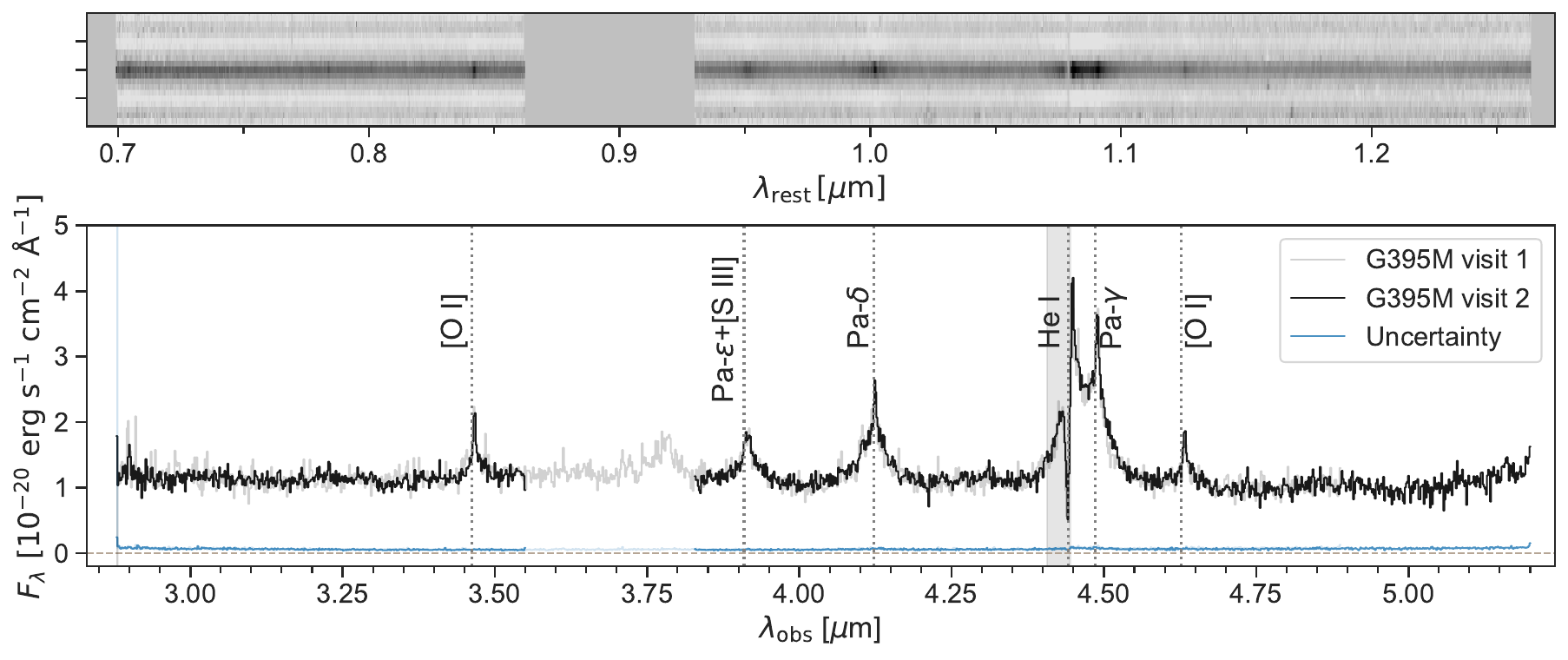}{0.95\textwidth}{(b)}
 } 
\caption{JWST/NIRSpec data of \rd. (a) The first row displays the 2D Prism spectrum, with the rest-frame wavelength shown along the $x$-axis; the histogram indicates the extracted region. The left panel in the second row shows the 1D spectrum in $F_\lambda$ as a function of the observed wavelength. Data and uncertainties are plotted in black and blue, respectively. The beige horizontal line is at $y=0$ to guide the eye. The cutout in the F444W band is shown to the right, with the slitlets overlaid.
(b) Same as the above figure set, but for the G395M visits. Only the 2D spectrum of the second visit is included, which has a higher signal-to-noise ratio.}
\label{fig:spec}
\end{figure*}

The NIRSpec/MSA low-resolution Prism spectra and the medium-resolution G395M spectra presented in this paper were obtained in two visits between January 16 and 19, 2024. For each visit, \rd\ was observed for 48\,min in the Prism/Clear mode and 48\,min in the G395M/F290LP mode, using a standard 3-shutter slitlet and 3-point nodding pattern.

The data from the different visits are reduced separately using \texttt{msaexp} \citep{Brammer2022}, corresponding to version 3 on the DAWN JWST Archive\footnote{\url{https://dawn-cph.github.io/dja}}. The details of the msaexp data reduction are presented in \citet{deGraaff2024:survey,Heintz2025}; notably, version 3 incorporates a major improvement to the absolute flux calibration, and reductions with local and global background subtraction strategies. The different versions do not impact the results of this paper, however. We use the local background subtraction throughout for consistency, because only local background subtracted spectra (obtained from image differences between the three nodded exposures) are available for the G395M mode.

The reduced Prism spectrum from visit 1 and G395M spectra from visits 1 and 2 are shown in Figure~\ref{fig:spec}. Due to the source location on the mask, the Prism spectrum from visit 2 falls largely ($\sim50\%$ of the trace) in the chip gap between the two NIRSpec detectors. We find that the wavelengths of the emission lines in this spectrum differ significantly from the Prism spectrum of visit 1 and also from the two grating spectra (by $\Delta v\sim 2000~\kms$), and that the flux is a factor 0.75 lower than measured in the other spectra. This likely reflects calibration issues for sources near the edge of the NIRSpec detectors, and in the remainder of the paper we therefore only use the Prism spectrum from visit 1.

To account for the wavelength-dependent slit losses we rescale the Prism spectrum to the NIRCam photometry by fitting a polynomial correction factor within our SED modeling (\S\ref{subsec:prosp:basic}). We then use the best-fit polynomial for the Prism spectrum to perform a slit loss correction to the G395M data.

\section{Spectral modeling\label{sec:data_analyses}}

Data for \rd\ are rich in features, with multiple emission lines, an absorption feature in \he, and a continuum sampled out to rest-frame 4~$\mu$m with MIRI. To extract the maximal amount of information, we approach the modeling from three perspectives, detailed in the following subsections. We note that each approach is tailored to a specific purpose: SED modeling for disentangling AGN/galaxy contributions and inferring stellar population properties, dedicated emission and absorption line fitting for gas kinematics, and photoionization modeling to study outflows. Together, these methods aim to provide the most comprehensive view of an LRD to date.

\subsection{Spectral Energy Distribution\label{sec:prosp}}

\subsubsection{Basic Set-up\label{subsec:prosp:basic}}

The available JWST and HST photometric data are jointly fitted with the full NIRSpec/Prism spectrum within the \prospector\ inference framework \citep{Johnson2021}, and the posteriors are sampled using the dynamic nested sampler \texttt{dynesty} \citep{Speagle2020}. The joint spectrophotometric fit follows the methodology in \citet{Wang2023:z12}. Here we reiterate the common elements for completeness, before detailing the modifications driven by the peculiar nature of \rd. Fitting the G395M spectrum simultaneously requires special treatment for the different line spread functions (LSFs) and slit losses, and we leave this for future work.

We convolve the model spectra with the NIRSpec/Prism instrumental resolution curve. We note that the instrumental broadening of NIRSpec for a point source has currently not yet been calibrated empirically. Given the compactness of \rd, we construct a model LSF using \texttt{msafit} \citep{deGraaff2024:jades} for a point source morphology, which at $4~\micron$ is approximately a factor 1.7 narrower than the LSF of a uniformly illuminated slit from the JDox\footnote{\url{https://jwst-docs.stsci.edu}}.
The resulting resolution from our model is higher, because it accounts for the fact that the spatial extent of the point source in the dispersion direction is narrower than the shutter width.
Additionally, to account for the wavelength-dependent slit losses in the Prism spectrum, we fit for a polynomial calibration vector of order 7 after applying a wavelength-independent calibration to scale the normalization of the spectrum to the photometry.

\subsubsection{Stellar Populations\label{subsec:prosp:gal}}

To motivate the need for a composite galaxy and AGN model, we start fitting \rd\ with a standard \prospector\ galaxy-only model. Here, no continuum emission from the AGN accretion disk is included, providing a useful baseline against which we can compare the results from our new composite model.

We adopt the MIST stellar isochrones \citep{Choi2016,Dotter2016} and MILES stellar spectral library \citep{Sanchez-Blazquez2006} in FSPS \citep{Conroy2010}. The star formation history (SFH) is described by the non-parametric \prospector-$\alpha$ model via mass formed in 7 logarithmically-spaced time bins \citep{Leja2017}, with the first bin width decreased to 13.47 Myr to better accommodate the younger age of the universe at $z>3$ \citep{Wang2024:sps}. A mass function prior and a dynamic SFH($M, z$) prior designed for deep JWST surveys are used as well \citep{Wang2023:pbeta}. Dust emission is modeled following \citet{Draine2007}, with uniform priors assumed on all parameters. The particular choice of uninformative priors is reflective of the fact that we are exploring a potentially exotic parameter space in dust emission models.
We refer the reader to \citet{Wang2024:sps} for a description of the rest of the parameters describing the stellar populations.

\subsubsection{AGN Accretion Disk\label{subsec:prosp:agn}}

The observation of unambiguous broad lines in the spectra of \rd\ motivates the modeling of the accretion disk continuum emission.
In brief, we approximate the direct UV/optical emission from an AGN accretion disk as piece-wise power laws following \citet{Temple2021}. This model is built from SDSS+UKIDSS+WISE AGN and spans the color-luminosity space probed by these surveys. The resulting spectrum of the accretion disk is similar to the SDSS composite spectrum from \citet{VandenBerk2001}, but mitigates the issue of host galaxy contamination.
The normalization of the AGN continuum is a free parameter,  parameterized as the ratio of the AGN to stellar flux densities at rest-frame 5500~\AA. The power-law slopes are fixed to the best fit values in \citet{Temple2021}. Since our wavelength range of interest probes the Rayleigh-Jeans tail of the blackbody emission, the slopes are not expected to vary significantly across different blackbody temperatures.

As for the dust attenuation, the stellar populations are affected by the diffuse dust, described by a \citet{Calzetti2000} curve with a flexible power-law slope \citep{Noll2009}. A free parameter is assigned to the fraction of the starlight allowed to live outside the dust screen entirely, suggesting that some of the blue stars have blown holes in the dust or otherwise exist outside it. This is similar to the scenarios invoked to describe the blue rest-UV slopes in otherwise very red sub-millimeter galaxies (e.g., \citealt{Hainline2011}).
We assume that the light from the accretion disk experiences the same dust attenuation as the stars, but is additionally reddened by a separate dust attenuation curve modeled as a power law with varying normalization and index. In other words, since the intrinsic AGN spectrum from the accretion disk is blue, we take the red continuum of \rd\ as indicative of significant dust presence. The range of the power-law attenuation is set such that the flattest slope is Calzetti-like while the steepest slope approximates that of the Small Magellanic Cloud.

Finally, we include a model component that describes hot dust emission from the AGN torus, which typically starts to be observable beyond rest-frame 2 $\mu$m. We adopt the torus model implemented in FSPS \citep{Conroy2009,Leja2018}, which is based on a simplified CLUMPY model \citep{Nenkova2008b} with two free parameters: the normalization and dust optical depth in the mid-IR. In practice, this means that the torus always peaks at the same temperature.

We take the combination of the above AGN accretion disk and torus model and the galaxy model (\S\ref{subsec:prosp:gal}) as the fiducial model of this paper. The relevant parameters are listed in Table~\ref{tab:theta}, and recovery tests performed on simulations are supplemented in Appendix~\ref{app:mock}. Importantly, the unknown nature of LRDs means that other possibilities cannot be ruled out a-priori. We thus additionally fit the data with three alternative models, including fitting with AGN light only, without the new dust component, and fitting the data without the UV continuum. We use these other models to infer the systematic errors in our inferred parameters, and to propose discerning hypotheses for future testing. Details are presented in Appendix~\ref{app:models}.

\begin{deluxetable*}{p{0.1\textwidth} p{0.4\textwidth} p{0.4\textwidth}}
\tablecaption{AGN and Dust Model Parameters and Priors\label{tab:theta}}
\tablehead{
\colhead{Parameter} & \colhead{Description} & \colhead{Prior}
}
\startdata
$f_{\mathrm{AGN_,bbb}}$ & ratio between the AGN and the host galaxy's flux densities at rest-frame 5500~\AA, pre-attentuation & log uniform: $\mathrm{min}= 10^{
-5}$, $\mathrm{max}= 500$ \\
$f_{\mathrm{AGN, torus}}$ & ratio between the AGN luminosity in the mid-IR and the host galaxy's bolometric luminosity & log uniform: $\mathrm{min}=10^{
-5}$, $\mathrm{max}= 3$ \\
$\tau_{\mathrm{AGN, torus}}$ & optical depth of AGN torus dust & log uniform: $\mathrm{min}=5$, $\mathrm{max}=150$ \\
$n_{\mathrm{dust, 2}}$ & power law index for a \citet{Calzetti2000} attenuation curve & uniform: $\mathrm{min}=-1.0$, $\mathrm{max}=0.4$ \\
$\hat{\tau}_{\rm dust, 2}$ & optical depth of diffuse dust seen by both the host galaxy and the AGN & uniform: $\mathrm{min}=0$, $\mathrm{max}=4$\\
$n_{\mathrm{dust, 4}}$ & power law index for the AGN-only attenuation curve & uniform: $\mathrm{min}=-1.8$, $\mathrm{max}=-0.8$ \\
$\hat{\tau}_{\rm dust, 4}$ & optical depth of dust seen only by the AGN & uniform: $\mathrm{min}=0$, $\mathrm{max}=4$\\
$f_{\rm nodust}$ & fraction of starlight that is not attenuated by the diffuse dust & uniform: $\mathrm{min}=0$, $\mathrm{max}=1$\\
\enddata
\tablecomments{The remaining \prospector\ parameters and priors adopted for the stellar population inference are described in \citet{Wang2024:sps}.}
\end{deluxetable*}

\subsubsection{AGN Torus\label{subsec:prosp:torus}}

In the above fiducial model, the accretion disk and the torus emissions are independent of each other, meaning that energy balance is not enforced. In other words, the fiducial model does not force the absorbed AGN energy in the rest-UV and optical to be re-emitted into the IR (though energy balance is enforced for the stellar light). This choice is made in light of the concern that the assumption of the attenuated luminosity equaling the emitted luminosity in the IR does not hold in the presence of an angular dependence of dust emission and absorption.
Dust emission in a complicated AGN dust torus geometry may have significant inclination dependence, which cannot be modeled with a single unresolved object (e.g., \citealt{Nenkova2008a,Nenkova2008b}). 
However, given the existing IR upper limits, we find it valuable to calculate the energy balance. We thus consider two variations on the fiducial model.

First, we ignore the IR limits, and let the total AGN luminosity absorbed by dust in the fiducial model to be re-emitted in the IR. In other words, we scale our simplified CLUMPY torus model spectrum by the luminosity absorbed by the dust. The resulting torus emission is then compared to the IR photometric observations.

Second, we perform the test more broadly on the full CLUMPY model \citep{Nenkova2008b}. As mentioned earlier, the torus model as implemented in \prospector\ is a simplified version. A wide range of factors can affect the resulting SED of the dust emission; accordingly the CLUMPY model has 6 parameters producing a wide range of observed SED shapes, and other models have a different physical picture entirely with their own associated SED shapes \citep{Dullemond2005,Honig2006}. 
A detailed characterization of the torus model is out of the scope of this paper. We conduct a simple experiment by randomly drawing CLUMPY models and scaling them to match the total AGN luminosity attenuated by dust. This test allows more freedom in interpreting the IR limits, including in particular extending the range of temperatures in the dust to colder values and allowing variation in the viewing angle.

\subsubsection{Emission-line Marginalization\label{subsec:prosp:elines}}

In the \prospector\ fit nebular and AGN emission lines are not interpreted using a physical model, but instead described by simple Gaussians using the methodology in \citet{Johnson2021}.  However, the standard method of using a single Gaussian is insufficient for our case, given the clear broad-line component visible in the observed spectra (Figure~\ref{fig:spec}). We thus implement a two-component Gaussian model for most lines, consisting of a narrow and a broad component. We also treat the velocities of the hydrogen lines separately from the metal lines, given the expectation that their line widths can differ from the other lines. This is mostly to ensure sufficient flexibility to accurately describe the lines. A total of four parameters are used to fit the emission lines, i.e., separate velocities for hydrogen lines compared to other lines with a broad and narrow component for each type.

The potential non-Gaussianity of the lines, some of which is visible in Figure~\ref{fig:spec}, is not accounted for. In order to avoid the likelihood being skewed by the residuals from the non-Gaussian line kinematics, as well as the potential systematic uncertainties in the spectral reduction, we impose a 5\% error floor in the spectroscopic data, same as the error floor usually applied for photometry. We also include a multiplicative noise inflation term as a free parameter, with a prior range from 0.5 to 5, serving as both a pressure valve in the event of significant model mismatch and as an additional check on the quality of the fit.

\subsection{Emission-line Decomposition\label{sec:line_fitting}}

Separate from the fitting with \prospector, we perform kinematic modeling of the brightest broad lines to measure the widths and luminosities of the broad and narrow line components. This separate fit is done to have more fine-tuned understanding of the line kinematics outside of the \prospector\ fit. We use the medium-resolution spectra covering $2.9-5.1~\micron$ to measure the properties of the \pad, \pag\ and \he\ lines, and fit this complex simultaneously to account for blended lines. 

We model each emission line with a broad and narrow Gaussian line profile, and allow for a velocity offset between the two components. We assume uniform priors for the redshift, line fluxes, and velocity dispersions ($\rm \sigma_{\rm broad}\in[700,5000]~\kms$; $\rm \sigma_{\rm narrow}\in[0,700]~\kms$). The local continuum is modeled as a 2nd-order polynomial. In addition to this emission model, we find a strong absorption feature at $\approx4.2~\micron$ present in both medium-resolution spectra. This is clearly visible in Figure~\ref{fig:spec}, and a zoom-in is shown in Figure~\ref{fig:blr_fit}. We model this feature as a blue-shifted outflow in the \he\,$\lambda$\,10830\,\AA\ line, and assume a Gaussian line profile with uniform priors for the flux, velocity dispersion ($\rm \sigma_{\rm outflow}\in[0,1000]~\kms$) and velocity offset with respect to the narrow \he\ component $\Delta v \in[0,2000]~\kms$. 

As mentioned in Section~\ref{subsec:prosp:basic}, we use a model LSF for a point source morphology \citep{deGraaff2024:jades}. Although the model LSF has a systematic uncertainty of $\sim10-20\%$, we find in Section~\ref{subsec:lines} that the narrow line component is substantially broader than the LSF, and is therefore unaffected. 
Furthermore, to robustly fit the narrow lines in our data we need to account for the under-sampling of the LSF ($\rm FWHM_{LSF}\sim1~pix$) and narrow emission lines. To do so, we construct our model on a wavelength grid that is over-sampled by a factor 5. After convolution with the LSF, the model is integrated and sampled at the original pixel resolution.

We leverage the data from the two independent visits for a better sampling of the LSF. Instead of combining the two spectra, which may introduce further correlated noise, we simultaneously fit to the two spectra. We use the Markov Chain Monte Carlo (MCMC) ensemble sampling method implemented in \texttt{emcee} \citep{emcee} to estimate the posterior distributions of the model parameters.

In addition to our modeling of the medium-resolution data, we also model the low-resolution Prism data, as we find that both the \Ha and \Pab lines are significantly broader than the Prism LSF. Due to significant issues in the wavelength and flux calibration in the prism spectrum from the 2nd visit, we only use the data from the first visit (i.e., consistent with the \prospector\ modeling). We fit the two lines individually, following a similar approach as for the medium-resolution data, but with some small modifications. 

The observed wavelength of \Pab falls on the edge of the spectrum, and the line is therefore partially cut off. With the added complication of the severe under-sampling of the LSF and uncertainty in the LSF, we adjust our model in two ways: (i) we assume that there is no velocity offset between the broad and narrow line, and (ii) we multiply the instrument broadening ($\rm \sigma_{LSF}$) by a nuisance parameter $f_{\rm LSF}$ to account explicitly for the uncertainty in the LSF. We assume a truncated Gaussian prior for the latter, with $f_{\rm LSF}\in[0.8,1.5]$, a mean of $\mu(f_{\rm LSF})=1.1$ and dispersion $\sigma(f_{\rm LSF})=0.2$ \citep[see Appendix A of][]{deGraaff2024:jades}.

To fit the \ha\ line we include the \Nii$_{6549,6585}$ doublet in the model. We assume that the \Nii\ lines are narrow and equal to the narrow line width of \ha, and fix the flux ratio of the doublet to 1:2.94. Because the narrow line width is a factor $>4$ narrower than the LSF, the width cannot be constrained from the Prism data alone. We therefore assume a Gaussian prior for the velocity dispersion of the narrow line based on the estimates from the well-resolved Paschen lines: $\mu(\rm \sigma_{narrow})=250~ \kms$ and $\sigma(\rm \sigma_{narrow})=50~ \kms$. 

Finally, as a consistency check, we perform the same double-Gaussian fit for the \ha\ line without including the \Nii$_{6549,6585}$ doublet. Additionally, we model both the \Pab\ and \ha\ lines using Lorentzian profiles to further validate our results.

\subsection{Single-epoch Black Hole Mass\label{sec:rmap}}

Reverberation mapping can obtain the radius of the broad-line region from the lag between the variability in the AGN continuum and the corresponding variability in the broad permitted lines (e.g., \citealt{Blandford1982}). Empirical correlations have since been derived between the radius and line luminosities and widths in the local universe (e.g., \citealt{Kaspi2000,Landt2013}). These relationships allow for the black hole mass to be estimated from single-epoch measurements.

Given the wavelength coverage of \rd, we estimate the black hole mass via three different scaling relations using the Balmer and Paschen series. Comparing these (mostly) independent estimators helps to mitigate the systematic uncertainties introduced by the application of these methods at higher redshifts and in different physical conditions than where they are calibrated. All line luminosities here are dereddened using the dust attenuation inferred from SED fitting. First, we use the H$\alpha$ scaling relation from \citet{Greene2005}:
\begin{multline}
	M_{\rm BH}/M_{\odot} = 2.0 \cdot 10^6 ~\cdot \\ \left( \frac{L_{{\rm H\alpha}}}{10^{42}~{\rm erg~s^{-1}}} \right)^{0.55}  \left(  \frac{ {\rm FWHM_{H\alpha}} }{10^3~{\rm km~s^{-1}}}  \right)^{2.06} ,
\end{multline}
which has an intrinsic systematic scatter of a factor of $\sim 2$.

Second, we adapt the \pab\ scaling relation from \citet{Kim2015}:
\begin{multline}
	M_{\rm BH}/M_{\odot} = 10^{7.04} ~\cdot \\ \left( \frac{L_{\rm Pa-\beta}}{10^{42}~{\rm erg~s^{-1}}} \right)^{0.48} \left(  \frac{ {\rm FWHM_{\rm Pa-\beta}} }{10^3~{\rm km~s^{-1}}}  \right)^{2} ,
\end{multline}
which has a lower intrinsic scatter of a factor of $\sim 0.2$ dex, but is inferred from a smaller reverberation mapping sample \citep{Kim2010}.

Third, we consider the scaling relation between rest-IR Paschen-series lines and the rest-IR continuum \citep{Landt2013}:
\begin{multline}
	\log M_{\rm BH} = 0.88 ~\cdot \\  \left( 2 \log \frac{ {\rm FWHM_{\rm H}} }{{\rm km~s^{-1}}} +  0.5 \log \frac{\nu L_{\nu, \rm 1 \mu m}}{{\rm erg~s^{-1}}} \right) - 17.39 .
\end{multline}
Although this relation has a high intrinsic scatter of $\approx1\,$dex, it allows us to utilize the better determined \pad\ and \pag\ lines, and serves as a consistency check.

In addition, we estimate the AGN bolometric luminosity in two ways to facilitate a comparison to the Eddington limit.
First, we apply a standard bolometric correction factor of 10 to the dust-corrected luminosity at rest 5100~\AA\ (e.g., \citealt{Richards2006,Shen2020}), and quote this value as our fiducial bolometric luminosity. While imperfect, this model-dependent value represents our best guess for the intrinsic luminosity of the AGN in the context of our preferred AGN model.
Second, we base our estimate on the inferred model spectra. We integrate over the observed spectrum with the galaxy contribution subtracted, and then add in the luminosity attenuated by dust. The latter is approximated as the inferred pre-attenuation AGN luminosity, minus the post-attenuation AGN luminosity, based on the dust content from SED fitting. This approximation is likely to be an underestimation, due to the unobserved peak (``big blue bump"; \citealt{Sanders1989}).

\subsection{Photoionization Modeling\label{sec:cloudy}}

We estimate the hydrogen column density associated with the \he\ absorber by running photoionization models using the C23 version of \cloudy\ \citep{Chatzikos2023}. We adopt the standard AGN radiation field which is a modified version of the \citet{Mathews1987} SED with a sub-millimeter break at 10 microns. We calculate a two-dimensional grid of models with hydrogen number density varied in the range \mbox{2 $< \log(n_{\textrm H})$[cm$^{-3}$] $<$ 10} in 0.5 dex steps and ionization parameter varied in the range \mbox{$-4.5 < \log(U) <$ 0} in 0.3 dex steps. The elemental abundances are fixed to solar. The models are run until they reach a hydrogen column density of \mbox{$\log N_{\textrm H}$[cm$^{-2}$] = 24}.

For each model, we compute $N_{\rm H}$ and $N_{\rm He}(2^3S)$ as a function of radius. We use 1D interpolation to find the radius where the predicted $N_{\rm He}(2^3S)$ most closely matches the measured value, and record the corresponding $N_{\rm H}$ at that radius. Models that do not reach the observed $N_{\rm He}(2^3S)$ are excluded from further analysis.

\section{Results\label{sec:res}}

Several previous studies have attempted to disentangle the AGN or galaxy origin of LRDs. \citet{Barro2023} and \citet{Kocevski2023} suggested that the red continuum can be explained by either a heavily obscured quasar or a dusty starburst galaxy, whereas \citet{Labbe2023:agn} used ALMA non-detections to infer that the compact red sources are not typical dusty star-forming galaxies.
The flat MIRI colors have further led to attempts to model the continuum as a reddened but old stellar population \citep{Williams2023}. 

With those ideas in mind, we begin by reporting the size measurements in the long and short wavelengths (\S \ref{subsec:size}). These may shed light on the AGN or galaxy origins as \rd\ is better resolved owing to its lower redshift. We then proceed to present results from fitting only starlight to the rest UV/optical continuum of \rd\ (\S \ref{subsec:res:galonly}). This model serves as a useful benchmark, against which we evaluate our AGN and galaxy composite model (\S \ref{subsec:res:fid}). Finally, we explore the family of models that are consistent with the mid-to-far IR data and limits, particularly in the context of dusty torus models that characterize typical AGNs (\S \ref{subsec:ir}). Parameter constraints, including inferred properties of the host galaxy and the AGN are listed in Table~\ref{tab:sps}, and line fluxes and kinematics are listed in Table~\ref{tab:elines}.

\subsection{Sizes in the UV and Optical\label{subsec:size}}

\rd\ is unresolved in the long wavelength filters, corresponding to the rest-frame optical emission, with an inferred effective radius of $r_{\rm e} \le 0.3 \ {\rm pixel}\ (\approx 0.1~{\rm kpc})$. However, comparison of the 1D radial profile to the PSF profile as well as 2D S\'ersic profile fitting in the F115W filter, suggest that the source is marginally resolved in the rest-frame UV with $r_{\rm e} = 1.21 \pm 0.21\ {\rm pixel}\ (\approx 0.2~{\rm kpc})$. Details on the morphological modeling can be found in Appendix~\ref{app:morph}.

Critically, this analysis is independent of the SED modeling. The apparent wavelength-dependence of the sizes already suggests that the UV and rest-frame optical trace different physical components.

\begin{deluxetable}{lll}
\tablecaption{Observations\label{tab:obs}}
\tablehead{
\colhead{Basic measurement} & \colhead{}
}
\startdata
RA [deg] & 34.244201 \\
Dec [deg] & -5.245872 \\
$z_{\rm spec}$ & $3.1034 \pm 0.0002$ \\
\hline
\hline
Filter & Flux [AB mag] & Instrument\\
\hline
F435W & $29.10 \pm 1.70$ & HST/ACS\\
F606W & $27.45 \pm 0.21$ & HST/ACS\\
F814W & $27.19 \pm 0.13$ & HST/ACS\\
F090W & $27.18 \pm 0.17$ & JWST/NIRCam\\
F115W & $26.54 \pm 0.09$ & JWST/NIRCam\\
F125W & $26.33 \pm 0.15$ & HST/WFC3\\
F140W & $26.23 \pm 0.25$ & HST/WFC3\\
F150W & $25.71 \pm 0.05$ & JWST/NIRCam\\
F160W & $25.64 \pm 0.07$ & HST/WFC3\\
F200W & $23.74 \pm 0.05$ & JWST/NIRCam\\
F277W & $22.34 \pm 0.05$ & JWST/NIRCam\\
F356W & $22.33 \pm 0.05$ & JWST/NIRCam\\
F410M & $22.09 \pm 0.05$ & JWST/NIRCam\\
F444W & $21.96 \pm 0.05$ & JWST/NIRCam\\
F770W & $21.53 \pm 0.01$ & JWST/MIRI\\
F1800W & $21.12 \pm 0.05$ & JWST/MIRI\\
MIPS 24$\mu$m & $<20.21$\tablenotemark{\scriptsize{a}} & Spitzer\\
PACS 100  & $<15.96$\tablenotemark{\scriptsize{a}} & Herschel\\
PACS 160  & $<15.10$\tablenotemark{\scriptsize{a}} & Herschel\\
\enddata
\tablenotetext{a}{3$\sigma$ upper limit.}
\end{deluxetable}

\begin{deluxetable}{ll}
\tablecaption{Inferred Properties of the Host Galaxy and the AGN, and Ancillary Parameters\label{tab:sps}}
\tablehead{
\colhead{Model-inferred property\tablenotemark{\scriptsize{a}}} & \colhead{}
}
\startdata
log~$M_{*}/\msun$  & $9.43^{+0.10}_{-0.11}$ \\
Age [Gyr] & $0.69^{+0.13}_{-0.14}$ \\
SFR [$\msun\,{\rm yr}^{-1}$] & $1.55^{+1.04}_{-0.67}$ \\
log~sSFR [${\rm yr}^{-1}$] & $-9.22_{-0.27}^{ +0.24}$\\
log~$Z_{*}/{\rm Z}_\odot$\tablenotemark{\scriptsize{b}} & $-1.27^{+0.67}_{-0.48}$ \\
$\hat\tau_{\rm dust,2}$ & $0.21^{+0.14}_{-0.09}$ \\
$n_{\rm dust,2}$ & $-0.62^{+0.29}_{-0.26}$ \\
$f_{\rm no dust}$ & $0.19^{+0.12}_{-0.11}$ \\
$f_{\rm AGN, 7500A}$\tablenotemark{\scriptsize{c}} & $11.69^{+3.01}_{-3.87}$ \\
$f_{\rm AGN, torus}$ & $12.11^{+11.34}_{-5.55}$ \\
$\hat\tau_{\rm torus}$ & $14.08^{+12.19}_{-5.97}$ \\
$\hat\tau_{\rm dust,4}$ & $2.79^{+0.12}_{-0.14}$ \\
$n_{\rm dust,4}$ & $-1.75^{+0.05}_{-0.03}$ \\
$A_{\rm V, dust,2}$ & $0.22^{+0.15}_{-0.10}$ \\
$A_{\rm V, dust,4}$ & $3.03^{+0.13}_{-0.15}$ \\
\hline
\hline
Additional inferred property &  \\
\hline
$\log M_{\rm BH}/\msun$, H$\alpha$\tablenotemark{\scriptsize{d}} & $9.06\pm 0.03$; $8.94\pm 0.02$ ($\pm 0.3$)\tablenotemark{\scriptsize{e}}\\
$\log M_{\rm BH}/\msun$, Pa$\beta$ & $8.93\pm 0.08$; $8.86\pm 0.08$ ($\pm 0.2$)\tablenotemark{\scriptsize{e}}\\
$\log M_{\rm BH}/\msun$, Pa$\delta$\tablenotemark{\scriptsize{d}} & $7.81\pm 0.06$; $7.74\pm 0.06$ ($\pm 1.0$)\tablenotemark{\scriptsize{e}}\\
$\log M_{\rm BH}/\msun$, Pa$\gamma$\tablenotemark{\scriptsize{d}} & $7.67\pm 0.04$; $7.59\pm 0.04$ ($\pm 1.0$)\tablenotemark{\scriptsize{e}}\\
Bolometric luminosity [$\lsun$]\tablenotemark{\scriptsize{f}} & $(12 \pm 1) \times 10^{12}$; $(2.2 \pm 0.2) \times 10^{12}$ \\
Eddington luminosity [$\lsun$] & $(2 - 33) \times 10^{12}$ \\
\enddata
\tablenotetext{a}{Posterior moments from \prospector.}
\tablenotetext{b}{Stellar-phase metallicity.}
\tablenotetext{c}{Ratio between the AGN and the host galaxy's flux densities around rest-frame 7500~\AA, post-attenuation.}
\tablenotetext{d}{The black hole (BH) mass in the first column is estimated assuming the fiducial model, and that in the second column is based on an alternative dust model (see Appendix~\ref{app:models}).}
\tablenotetext{e}{The uncertainty in the parentheses indicates the intrinsic scatter in the scaling relationship used for estimating the BH mass.}
\tablenotetext{f}{The bolometric luminosity in the first column are from a standard bolometric correction to the luminosity at rest 5100\,\AA, and that in parentheses is estimated from the observed spectrum and the inferred dust attenuation.}
\end{deluxetable}

\subsection{Galaxy-only Best Fit\label{subsec:res:galonly}}

\begin{figure*} 
\gridline{
  \fig{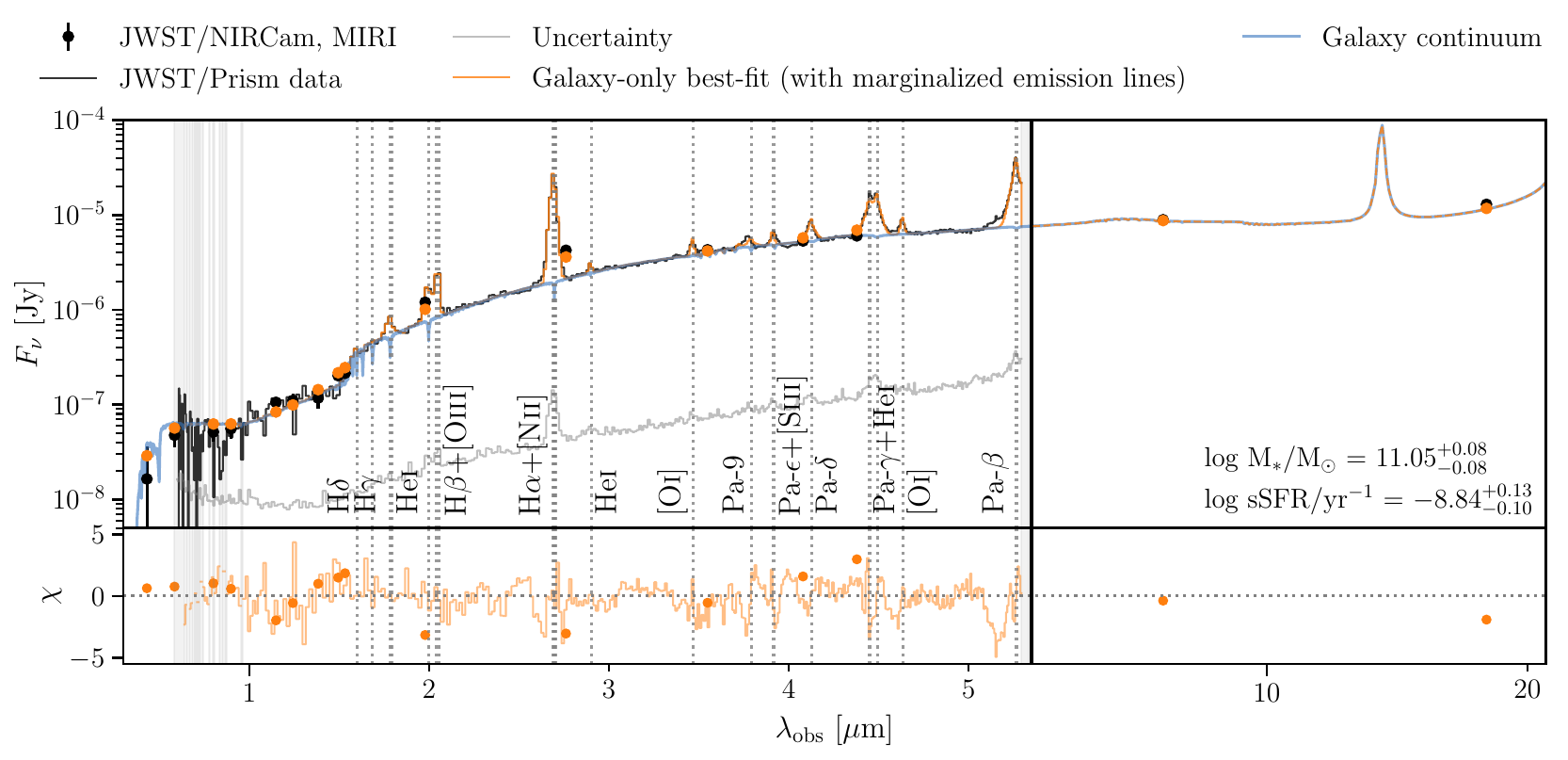}{0.95\textwidth}{}
}
\caption{Spectrophotometric modeling with starlight only. The photometric and spectroscopic data are shown in black, whereas the uncertainty in the observed spectrum is shown in gray. The best-fit model spectrum, which includes the marginalized emission lines) is plotted in orange, whereas the unconvolved galaxy model spectrum is over-plotted in blue. The $x$-axis transits from a linear to a log scale at 5.4~$\mu$m, indicated by the black vertical line, to show the MIRI detections. The spectral regions that are masked due to low S/N, blended, and incomplete lines are shaded in gray. The emission lines included in the marginalization are annotated. $\chi$ is included in the bottom.
}
\label{fig:sed_galonly}
\end{figure*}

As seen from Figure~\ref{fig:sed_galonly}, the galaxy-only model produces a good fit in the rest-frame UV-optical and an inferred star mass of $\sim 10^{11}~ \msun$ with an sSFR of $\sim 10^{-9}~ {\rm yr^{-1}}$. The rest of the inferred properties are listed in Appendix~\ref{app:models}, including its mass-weighted age of $\sim 0.8$~Gyr and $\Av$ of $\sim 3$.

It would be unusual but not impossible to have such a massive galaxy (e.g., the quiescent galaxy at $\zspec=3.97$, which has $\Av \sim 1.4$ in the center; \citealt{Setton2024}). Assuming the strong broad-line features (\ha\ rest-frame EW $\sim 800$~\AA) are of AGN origin, it would be surprising to have such a massive black hole with very little continuum contribution from the accretion disk, although AGN broad lines may still exist in older galaxies \citep{Carnall2023}.
Moreover, the small size of \rd\ (effective radius of $\approx 0.2$ kpc in the UV) implies a very high stellar density \citep{Baggen2023}. This size would also make \rd\ a small galaxy relative to its peers, but not an outlier among known old galaxies at similar mass and redshift \citep{Wright2023,Ji2024}.
However, the inferred narrow line widths of the \pag\ and \pad\ lines ($\sim 250\,\kms$) and small size imply a dynamical mass ${\rm M_{dyn}}\sim 5\,R_{\rm e}\sigma^2/G \sim 10^{10}\,\msun$ \citep[where $G$ is the gravitational constant; see e.g.,][]{Cappellari2006}, which would be difficult to reconcile with a stellar mass estimate of $10^{11}\,\msun$ even when considering projection effects that may underestimate the true velocity dispersion of the system.

Therefore, we favor the fiducial galaxy and AGN joint fit presented in the following section, but cannot definitively rule out a galaxy-dominated continuum. Model comparison statistics based on the Bayes factors \citep{Trotta2008} shows no strong preference for either model. Follow-up deep medium-resolution data to reveal possible stellar absorption lines, or deeper rest-frame $>30~\mu$m data, will help to discriminate the different scenarios.

\subsection{UV/Optical Best Fit\label{subsec:res:fid}}

\begin{figure*} 
\gridline{
  \fig{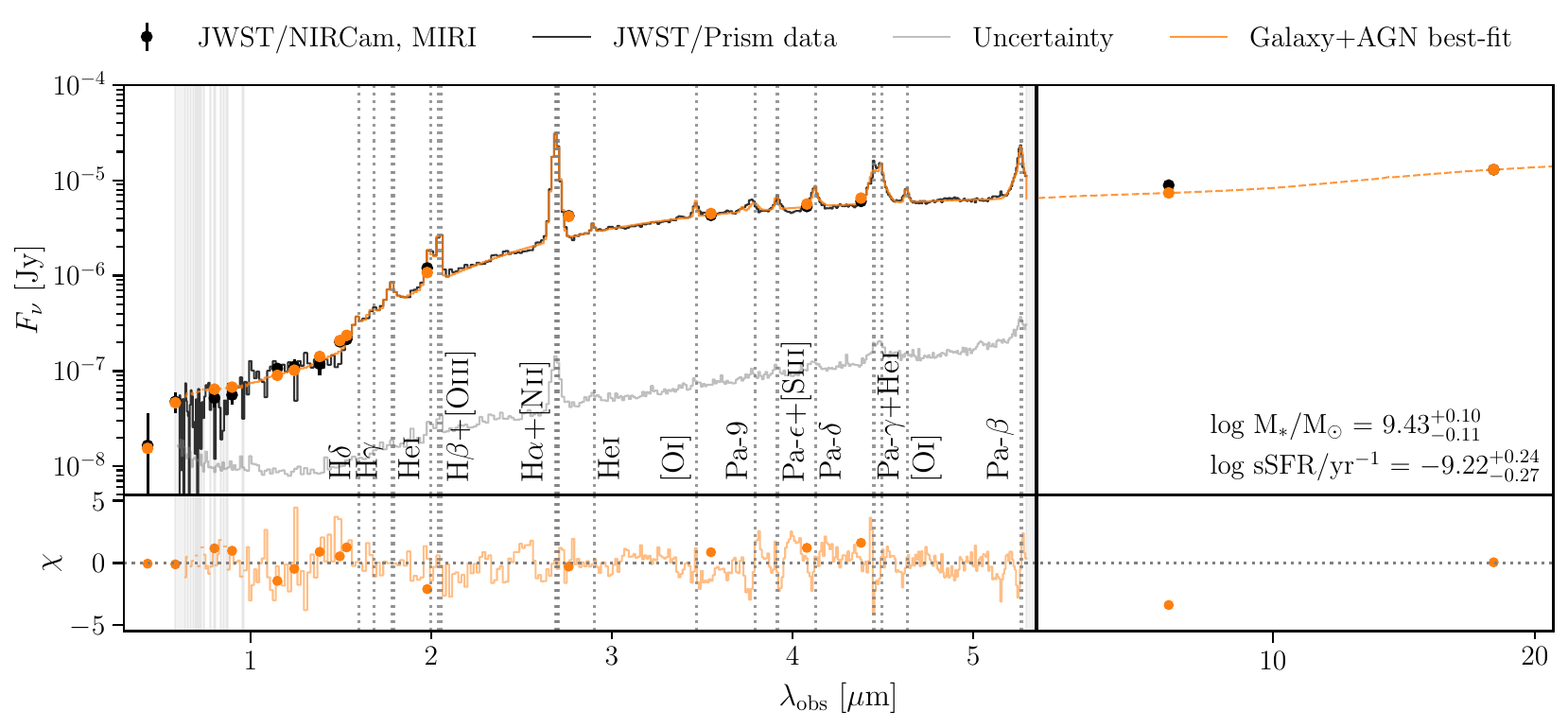}{0.9\textwidth}{(a)}
}
\gridline{
  \fig{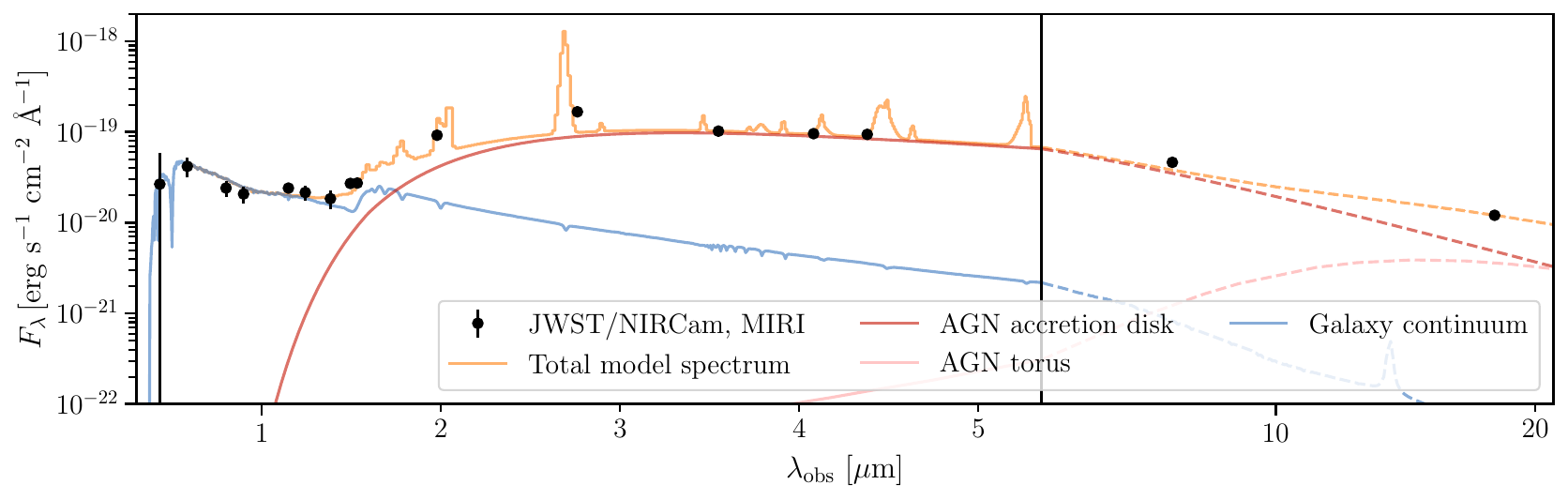}{0.9\textwidth}{(b)}
}
\gridline{
  \fig{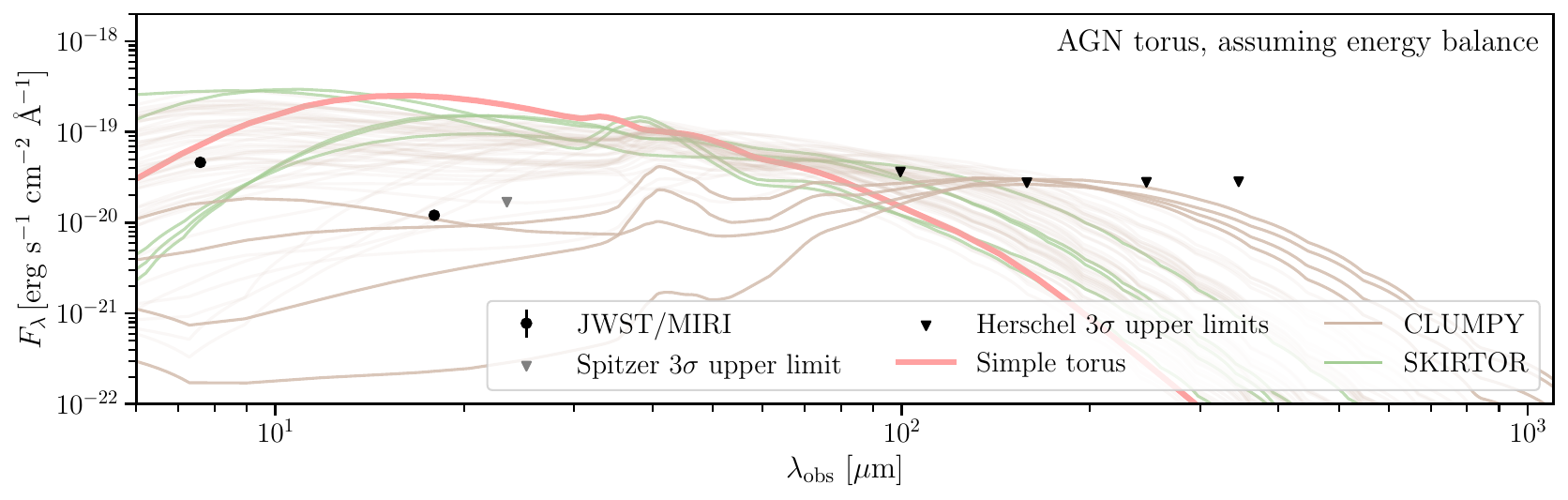}{0.9\textwidth}{(c)}
}
\caption{Spectrophotometric modeling with the fiducial AGN and host galaxy composite model. (a) Data and best-fits are shown in the same manner as Figure~\ref{fig:sed_galonly}. (b) The various components of the model spectrum are illustrated here. Without assuming energy balance, the model predicts the accretion disk emission dominating the spectrum at $\gtrsim 2~\mu$m in the observed frame, with little contribution from the torus. (c) Torus model spectra, assuming that the attenuated AGN luminosity in the UV/optical is re-emitted in the IR. The rescaled simple torus used in the \prospector\ fit is plotted in pink. The resulting torus emission violates the MIRI detection and the upper limit from Spitzer/MIPS. Over-plotted in brown are a set of torus models from CLUMPY \citep{Nenkova2008b} that can produce marginally consistent spectra with the IR data. No SKIRTOR model can result in a torus emission that is consistent with the data; we shown a few random draws in green for illustration.}
\label{fig:sed}
\end{figure*}

In Figure 4 we show our fiducial galaxy and AGN joint model. An immediate observation is the capability of our fiducial model to generate a very red slope in the rest optical region and a blue slope in the rest near-UV, providing an effective fit to the observed data. 
This v-shape in $F_\lambda$ is difficult to reproduce with a simple AGN-only model (see Appendix~\ref{app:models}).
As depicted in Figure~\ref{fig:sed}, starlight makes up the rest-UV, whereas the light from the reddened AGN accretion disk dominates the spectrum red-ward of rest 4000~\AA.

The model AGN continuum emission is very bright and red due to the large implied dust attenuation and dominates over the wavelength range probed by JWST/MIRI. The torus contributes little to the total light in this fiducial fit even at rest-frame 5~$\mu$m.

The host galaxy has a stellar mass of $\sim 10^9 ~ \msun$, and a mass-weighted age of $\sim 0.3$ Gyr ($\sim 15$\% of the age of the universe). It is metal-poor ($\sim 0.05 ~ {\rm Z}_\odot$), and slightly below the star-forming main sequence when comparing its sSFR of $\sim 10^{-8.7} ~ \msun\,{\rm yr}^{-1}$ to the value of $\rm sSFR(M_*\sim10^9 ~ \msun)\sim 10^{-8.3} ~ \msun\,{\rm yr}^{-1}$ as reported in \citet{Speagle2014}. The dust attenuation is $\Av\sim 0.5$, with $\sim 10$\% of the total stars outside the dust screen. The inferred star-formation history resembles that of a post-starburst galaxy, likely driven by the presence of a Balmer break.

The preference for a metal-poor solution is primarily caused by the need to predict the UV excess, since the fiducial model essentially couples the UV light to the galaxy. A natural question is thus whether the host galaxy properties are heavily influenced by the UV continuum. We address this question by fitting our fiducial model to the rest optical part of the spectrum only in Appendix~\ref{app:models}. In brief, we find that the inferred stellar mass as well as the inferred star-formation history in this case are similar to those from the fiducial model, suggesting that the UV continuum is not driving the inferred SFH or stellar mass. 

Notably, the above results from SED fitting are consistent with the size measurements: the emission in the rest-frame UV originates from the host galaxy and thus is extended spatially, whereas the rest-frame optical emission is dominated by the AGN that is point-like in nature. However, as noted in \citet{Wang2024:bb}, this interpretation requires a peculiar AGN continuum shape, which conspires with the host galaxy to produce the observed spectral break.

\subsection{IR Constraints and an Inability to Model the Mid-IR with a Standard Hot Torus\label{subsec:ir}}

While our fiducial model does a fine job fitting the rest-frame UV and optical constraints, the MIRI data are surprisingly flat compared to our expectations for a hot dust component from the torus \citep{Bosman2023}. Thus, we consider the alternative modeling assumptions as listed in Section~\ref{subsec:prosp:torus}, where we attempt to self-consistently model the reddened AGN disk emission and the re-emitted torus dust. The AGN torus spectrum from our fiducial \prospector\ model, re-scaled with the total energy absorbed by dust, is shown in Figure~\ref{fig:sed}\,(c). While the model spectrum is marginally consistent with the Herschel upper limits, the MIRI/F1800W and MIPS points are at least a factor of 20 below the predictions from the scaled torus model.

The inability of simple hot AGN torus dust models to explain the observations motivates a more full exploration of the CLUMPY model set \citep{Nenkova2008b}. We draw random models that obey our mid-to-far IR upper limits when scaled by the total energy absorbed by dust, and show 100 examples, also in Figure~\ref{fig:sed}\,(c). This shows that additional freedom in torus clump size, torus shape, and angular dependence of the dust emission and absorption can produce colder AGN dust models which are marginally consistent with the observational constraints.
These lower temperatures can be caused by thicker dust shielding which produces a wider and overall colder range of dust temperatures. There is a known population of hot-dust-deficient AGNs which offers further support for this scenario \citep{Hao2010,Jiang2010,Mor2011,Lyu2017}.

It is worth pointing out that due to the complex geometry and physics involved in modeling torus, other models exist. We thus repeat the exercise above with the SKIRTOR \citep{Stalevski2012,Stalevski2016} model set. A few model draws are included in Figure~\ref{fig:sed}\,(c). In this case, it is not possible to find a torus model spectrum that is consistent with the observed IR data points. It is not surprising that CLUMPY models can make colder dust, as there can be a lot more self-shielding; that is, the dust is clumpy, giving a colder tail to dust temperatures.

We note that since the AGN bolometric luminosity from $L_{5100}$ may be $\sim 6$ times higher than that inferred from the continuum model, we perform an additional test where we increase the remitted IR luminosity by a factor of 2. In this case, the full CLUMPY model struggles to produce a torus spectrum that is consistent with the IR limits, just making the challenge to model the torus more acute. 

Three physical scenarios can be invoked to explain the apparent lack of torus emission.
First, the dust emission from LRDs may be in a relatively unexplored regime, being colder than any model in the CLUMPY database. Second, a majority of the energy originated from the big blue bump escapes through some combination of anisotropic dust attenuation or hard X-ray photons and so is not absorbed by dust. Third, the UV part of the accretion disk could be suppressed in a super-Eddington flow \citep[e.g.,][]{Abramowicz1988,Kubota2019}. Deeper rest-frame mid/far-IR constraints would help to reveal the nature of the source.

\subsection{Gas Kinematics\label{subsec:lines}}

\begin{figure*} 
\gridline{
  \fig{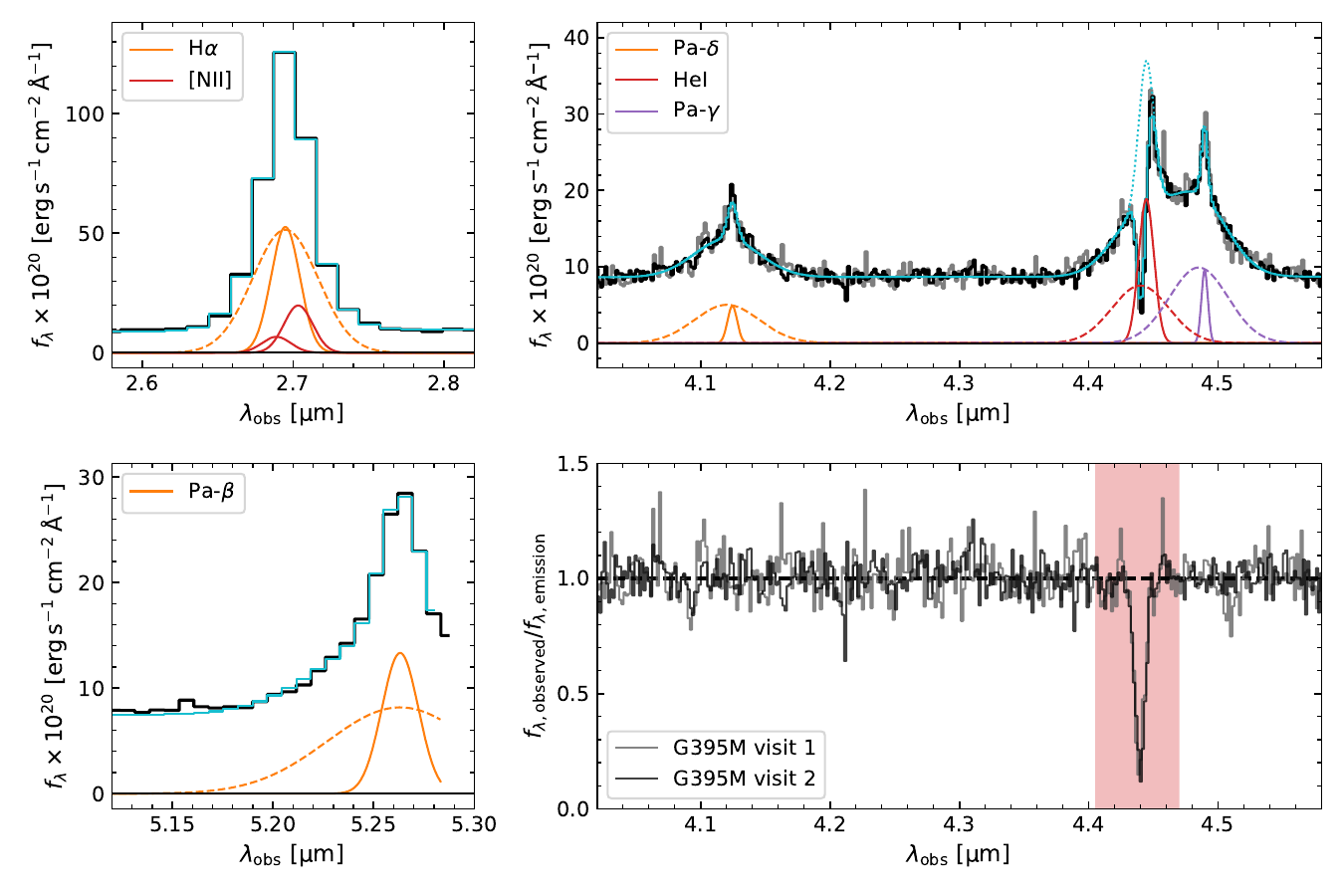}{0.95\textwidth}{}
} 
\caption{Kinematic emission line decomposition. On the left, we show the emission line fits to the Prism spectrum for the blended \ha\ and \nii\ complex (top) and \pab\ line (bottom). Although the narrow line widths are poorly constrained due to the broad LSF (and the partial coverage of \pab), the broad lines are well constrained. On the right, we show the results of our fitting for the \pad\ and blended \he\ and \pag\ lines in the G395M spectra (top). The total model that includes emission and absorption is shown as a solid curve in cyan, where as the total emission model, i.e., before any absorption, is shown as a dotted curve. A strong blue-shifted absorption feature in the \he\ line is found in the two independent G395M spectra, which we interpret as an ionized outflow. The bottom right shows the observed spectrum divided by the best-fit model emission spectrum. The red shaded region indicates the area used to integrate the optical depth and compute the column density of the absorbing gas. 
}
\label{fig:blr_fit}
\end{figure*}

\begin{deluxetable*}{lllllll}
\tablecaption{Results of Emission-line Decomposition\label{tab:elines}}
\tablehead{
\colhead{Emission} & \colhead{Flux (total)} & \colhead{Flux (narrow)} & \colhead{$\sigma$ (narrow)} & \colhead{Flux (broad)} & \colhead{$\sigma$ (broad)} & \colhead{EW (total)} \\
\colhead{} & \colhead{$\rm 10^{-18}~erg~s^{-1}~cm^{-2}$} & \colhead{$\rm 10^{-18}~erg~s^{-1}~cm^{-2}$} & \colhead{km~s$^{-1}$} & \colhead{$\rm 10^{-18}~erg~s^{-1}~cm^{-2}$} & \colhead{km~s$^{-1}$} & \colhead{rest-frame \AA}
}
\startdata
H$\alpha$ & $406^{+17}_{-14}$ & $126^{+23}_{-21}$ & $254^{+48}_{-48}$ & $280^{+18}_{-19}$ & $2180^{+62}_{-50}$ & $897^{+37}_{-32}$ \\
{[}N~$\textsc{ii}${]} & $63^{+14}_{-17}$ &  &  &  &  & $140^{+32}_{-37}$ \\
Pa-$\beta$ & $105^{+4}_{-4}$ & $30^{+5}_{-4}$ & $437^{+64}_{-65}$ & $74^{+4}_{-5}$ & $2048^{+213}_{-173}$ & $21^{+1}_{-1}$ \\
Pa-$\delta$ & $36^{+1}_{-1}$ & $5^{+1}_{-1}$ & $261^{+69}_{-57}$ & $32^{+1}_{-1}$ & $1824^{+120}_{-102}$ & $88^{+3}_{-3}$ \\
Pa-$\gamma$ & $62^{+3}_{-2}$ & $6^{+1}_{-1}$ & $153^{+18}_{-16}$ & $56^{+3}_{-2}$ & $1516^{+56}_{-56}$ & $174^{+7}_{-7}$ \\
He $\textsc{i}$ & $70^{+4}_{-3}$ & $28^{+2}_{-2}$ & $393^{+39}_{-38}$ & $42^{+3}_{-3}$ & $1492^{+77}_{-74}$ & $191^{+11}_{-10}$ \\
\hline
\hline
 Absorption \hspace{0.1em} & Outflow flux & ${\rm d}v_{\rm outflow}$ & $\sigma_{\rm outflow}$ & & & EW\\
 & $\rm 10^{-18}~erg~s^{-1}~cm^{-2}$ & km~s$^{-1}$ & km~s$^{-1}$ &  &  & rest-frame \AA \\
\hline
He $\textsc{i}$ & $25^{+2}_{-3}$ & $-231^{+24}_{-20}$ & $234^{+9}_{-8}$ &  &  & $-69^{+6}_{-8}$
\enddata
\end{deluxetable*}

We show the decomposition into narrow and broad components (\S\ref{sec:line_fitting}) in the Paschen and \Ha lines in Figure~\ref{fig:blr_fit}, and list the fluxes and velocity dispersions in Table~\ref{tab:elines}. We find that the narrow and broad line widths of the \pag\ and \pad\ lines are in good agreement, and that the narrow line flux comprises 13\% and 10\% of the total line flux, respectively. The narrow and broad line widths of the \pab\ emission are broader than the other Paschen lines by approximately $2\sigma$, and with a higher narrow-to-total flux ratio (29\%). Although some physical differences may be expected if the different line transitions trace gas in different parts of the broad line region (BLR), the \pab\ line is at the very edge of the spectrum which may bias the kinematic fitting. The \ha\ broad line is broader than the \pad\ and \pag\ broad lines by a factor $\approx 1.2$ and $1.4$, respectively; the narrow line flux is approximately 30\% of the total \ha\ line flux. This may imply that the \ha\ line traces a closer-in region of the BLR compared to the Paschen lines \citep[see e.g.,][]{Kim2010}. 

It is worth noting that as the narrow \ha\ line and the \nii\ lines are unresolved at the resolution of the Prism, the kinematic fit depends on the model assumptions made (at the $\approx 10\%$ level for the broad line width). When excluding the \nii\ lines in the double-Gaussian fit, the broad FWHM and flux are fully consistent with the fit that includes \nii. However, when using a Lorentzian profile, while the fit is equally good, the FWHM of the broad \ha\ line becomes $\sim 1800~\kms$, leading to a factor of 2.5 difference in width comparing to the double-Gaussian fit. In contrast, the Lorentzian fit applied to the \pab\ line yields more consistent results, with a FWHM that is $\sim 25$\% lower, but flux roughly twice as high, resulting in a similar black hole mass estimate. Therefore, although the Prism resolution is too low to robustly constrain the black hole mass from \ha, the various Paschen measurements still allow for order-of-magnitude arguments regarding the black hole mass (\S\ref{subsec:bhmass}).

Follow-up spectroscopy at higher resolution will be crucial to perform more careful modeling of the \ha\ and \nii\ emission line complex and robustly constrain the narrow-to-total line flux ratio, and to confirm the difference in the broad line width between the Balmer and Paschen lines. 

Interestingly, we find a blue-shifted absorption feature in the wings of the \he\,$\lambda$\,1.083\,$\mu$m line.
The \he\ absorption of \rd\ shows up in both G395M spectra, ruling out the possibility of a detector defect. 
We measure a velocity dispersion of $\sim 234~{\rm km ~s^{-1}}$, and a velocity offset of $\sim -231~{\rm km ~s^{-1}}$ with respect to the narrow He~\textsc{i} emission. The \he\ line is associated with the 2$^3$S metastable state, which is 19.7 eV above the ground state and is primarily populated by recombination from He~\textsc{ii}. Therefore, the absorption line effectively traces ionized gas. This makes it likely that the absorption tracers an outflow from \rd\ rather than intervening gas along the line of sight. 

A few other LRDs show absorption features, albeit in the \ha\ line. Whereas \citealt{Matthee2023:lrd} interpreted this as an outflow feature, \citet{Maiolino2023} proposed a dual AGN explanation. The addition of \rd\ suggests that such outflows may be a common feature among LRDs, although follow-up high-resolution observations of the Balmer lines will be needed to confirm whether the outflow is present in both \he\ and \ha.
Similar outflows in the \he\ line have been observed in multiple nearby AGN \citep[e.g.,][]{Leighly2011,Zhang2017,Pan2019}. However, we note that it is also possible that the \he\ line profile originates from Wolf-Rayet stars, consistent with the modest velocities \citep{Eenens1994,Stevens1999}. In this case, the P-Cygni profile is to be interpreted as a stellar wind rather than a galactic wind. Constraints on other possible Wolf-Rayet features will be needed to conclusively rule out the stellar wind scenario.

We estimate the optical depth of the \he\ absorption by constructing the emission-only spectrum from our kinematic modeling ($f_{\rm\lambda,emission}$), and integrating over the absorption line (red shaded region in Figure~\ref{fig:blr_fit}; e.g., \citealt{Savage1991}):
\begin{equation}
    \tau = \int \ln\left( \frac{f_{\rm\lambda,obs}}{f_{\rm\lambda,emission}} \right) {\rm d}\lambda\,.
\end{equation}
The column density, assuming that the source is fully covered by the absorber, is then computed as
\begin{equation}
    N_{\rm He}(2^3S) = \frac{m_{\rm e} c}{\pi e^2 f_0 \lambda_0} \tau\,,
\end{equation}
where $m_{\rm e}$ and $e$ are the electron mass and charge, $c$ is the speed of light, and $f_0$ and $\lambda_0$ are the wavelength and oscillator strength of the line transition. We find $N_{\rm He}(2^3S)=5.9\pm0.3\times10^{13}~{\rm cm}^{-2}$, where the uncertainty reflects both the uncertainty in the emission-only model and the observed flux density. This measured column density agrees well with measurements of \he\ absorbers found in nearby AGN \citep[e.g.,][]{Zhang2017,Pan2019}

\subsection{Mass Outflow Rate}

We estimate the mass outflow rate from this source using the outputs of the \cloudy\ modeling. The time-averaged mass outflow rate of a spherically symmetric, mass-conserving wind of finite radius is given by
\begin{equation}
\dot{M}_{\rm out} = 4 \pi C_\Omega \mu m_p N_H R_{\rm out} v_{\rm out},
\label{eq:mout}
\end{equation}
where $\mu$ is the mean atomic mass per proton (1.4), 4$\pi C_\Omega$ is solid angle subtended by the outflowing material, $R_{\rm out}$ is the outflow size and $v_{\rm out}$ is the outflow velocity \citep[e.g.,][]{Rupke2005}. We assume that the absorbing material covers 50\% of the solid sphere (i.e., $C_\Omega$ = 0.5), motivated by the fact that at least 50\% of quasars at \mbox{$z\sim$~2~--~4} show associated narrow absorption lines \citep[e.g.,][]{Misawa2007}.

The hydrogen column density and outflow size cannot be measured directly. However, we can constrain the range of reasonable values through the photoionization modeling. In Section~\ref{subsec:lines}, we measured a column density of \mbox{$\log N_{\rm He}(2^3S)/{\rm cm^{-2}} \simeq$ 13.8}.

\begin{figure*}[t]
\centering
\includegraphics[clip = True, trim = 0 100 0 0]{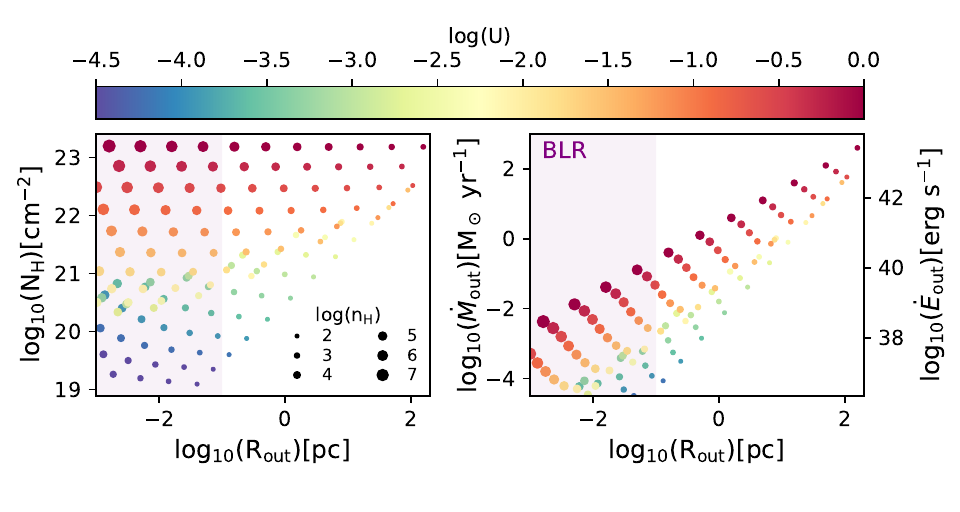}
\caption{Photoionization modeling with \cloudy. (Left) Inferred possible hydrogen column density of the \he\ absorber is plotted as a function of size. Each point represents a single model, where the color and size indicate the density and ionization parameter of that model, respectively. The purple shaded region covers sizes $<$~0.1~pc which would approximately represent an origin within the AGN broad line region. (Right) Inferred mass outflow rates are shown in the same manner. The possible mass outflow rates span 6 orders of magnitude, primarily due to the unknown gas density.}
\label{fig:outflow}
\end{figure*}

The left panel of Figure \ref{fig:outflow} shows the radius and $N_H$ of the \cloudy\ models when they reach the measured $N_{\rm He}(2^3S)$. The points are color-coded by the ionization parameter and the symbol sizes indicate the gas density. The observed $N_{\rm He}(2^3S)$ can be matched by absorbers at a wide range of radii and column densities. Rather than choosing representative values to calculate the mass outflow rate, we compute the mass outflow rate for each grid point individually. The right-hand panel shows the inferred mass outflow rate as a function of outflow size, as well as the corresponding kinetic luminosity of the outflow, given by $1/2~\dot{M}_{\rm out}v_{\rm out}^2$. 

There is a strong positive correlation between the radius and the mass outflow rate.
The lowest density models reach radii of $10-100$~pc with corresponding mass outflow rates of $1-200$~$\msun \, {\rm yr^{-1}}$. On the other hand, the highest density models imply radii of \mbox{$<$~0.1 pc}, which would place the absorber within the AGN broad line region, with corresponding mass outflow rates \mbox{$<$ 0.01~M$_\odot$ yr$^{-1}$}. The strong correlation arises because the mass outflow rate depends linearly on the outflow size (Equation~\ref{eq:mout}) and the model grid covers 5 orders of magnitude in outflow size. Conversely, the hydrogen column density is driven entirely by the ionization parameter and is virtually insensitive to hydrogen volume density (also shown in Figure 10 of \citealt{Ji2015}). The consequence is that the hydrogen column density does not show strong dependence on outflow size. To constrain the mass outflow rate better would require another estimate of the absorber size, more absorption lines, or both. 

\subsection{Black Hole Mass\label{subsec:bhmass}}

The various scaling relations consistently result in black hole masses in the range $\sim 10^{7.7} - 10^{9.1}~\msun$. This black hole mass is lower than typical UV-selected quasars at $z \sim 3$ \citep[e.g.,][]{Shen2012}, but, as it is roughly $2-42$\% of the stellar mass, similar to the unexpected large black hole to stellar mass ratios found at higher redshifts \citep{Goulding2023,Maiolino2023}. 

If we adopt the narrow line widths as $\sigma_{\rm gas} = 200-270$~km/s, then we find an expected range of $M_{\rm BH} = 4 \times 10^7 - 2 \times 10^8~\msun$. These values are consistent with our broad-line scalings, suggesting that while the stellar mass may be catching up, the black hole may still scale with the stellar velocity dispersion \citep{Chisholm2003,Maiolino2023}. 

Given the range of black hole masses, we obtain an Eddington luminosity of $\sim 2-12 \times 10^{12}~\lsun$, implying the black hole radiates at $\sim 20$\% of the Eddington limit to potentially super-Eddington.
However, we caution that systematic uncertainties, including the intrinsic scatter in the scaling relations and different dust models, which are not accounted for, likely dominate the error budget at $\gtrsim 1$~dex level for the above estimations.

\subsection{X-ray Non-detection}

\rd\ was observed by the Chandra X-ray Observatory in Cycle 17, as presented by \citet{Kocevski2018}. We adopt the exposure maps, event rates, and PSF models from those authors. At the position of \rd, the effective exposure time is $\approx170$~ks using the Advanced CCD Imaging Spectrometer-I. Assuming a standard power-law slope of $\Gamma = 1.9$ at $z=3$ \citep{Wang2021} and foreground galactic absorption $N_{\rm H} = 2.01 \times 10^{20} ~{\rm cm}^{-2}$, we directly measure the rest-frame $2-10$~keV counts at the position of the source. The resulting $3 \sigma$ upper limit on the X-ray luminosity is $<3 \times 10^{43} ~{\rm erg \,  s^{-1} }$, making the source nearly 100 times fainter than we would naively expect from the dereddened $L_{5100}$ luminosity of $\sim 10^{46} ~ {\rm erg \,  s^{-1} }$. This surprising result is discussed further in Section \ref{sec:discussion}.

\section{Discussion and Conclusions\label{sec:concl}}

In this paper, we present spectroscopic observations of a bright IR-luminous broad-line LRD using NIRSpec/Prism and G395M onboard JWST. We perform spectrophotometric modeling to infer the properties of the host galaxy and the AGN, and decompose the emission and absorption complexes to constrain the gas kinematics and measure the black hole mass. Additionally, we take a first look at the mass outflow rate as implied from the \he\ $\lambda$ 1.083$\mu$m absorption. Below we discuss the key findings in this work.

\subsection{A Spectroscopically Confirmed Broad-line AGN with a Deficit of Hot Dust}
\label{sec:discussion}

\rd\ is an unusually red and bright object compared to the parent RUBIES sample. In this paper, we argue that the bright and red rest-frame optical continuum is likely dominated by AGN light. The presence of an AGN is supported by the broad and symmetric Balmer and Paschen emission lines, strongly suggesting that we are seeing the AGN disk and broad-line region directly (i.e., an unobscured, but highly reddened AGN). In combination with the lack of broadened forbidden lines, the data disfavor an outflow interpretation. 
% The AGN-like nature is also supported by high-ionization lines detected in the mid-infrared (if we want to talk about this). 

We have developed a novel method to model the host galaxy and AGN within the \prospector\ Bayesian inference framework \citep{Johnson2021}. Our preferred model is a composite AGN and host model, in which the UV and the Balmer break are explained by galaxy light, and the AGN continuum from the outer accretion disk dominates in the optical and mid-IR. 
These results are further corroborated by our morphological analysis of \rd, in which we find the source to be marginally resolved in the rest-frame UV, but unresolved at rest-frame optical wavelengths (Appendix~\ref{app:morph}).

Explaining the rest-frame optical continuum light with pure galaxy continuum is possible, but yields a very high mass of $\sim 10^{11}~\msun$. While unresolved in the rest-frame optical, its measured half-light radius ($\sim 0.19$~kpc at F115W) would yield an extreme stellar surface density $\sim 10^{12}$ $\msun$\,kpc$^{-2}$, at least an order of magnitude higher than those of the massive, compact quiescent galaxies \citep{Bezanson2009}. Such an extreme stellar surface density would be in some tension with theoretical density limits from stellar feedback \citep{Hopkins2010,Grudic2019}, and is inconsistent with the dynamical mass inferred from the narrow \pag\ and \pad\ emission lines. Further, a galaxy interpretation also would lead to a very high \ha\ rest-frame EW of $\sim 800$~\AA, which is inconsistent with the inferred older stellar populations and also has not been seen in known AGNs \citep{VandenBerk2001,Stern2012}. 
Likewise, fitting a pure AGN model fails to reproduce the spectral shape around the Balmer break region (Appendix~\ref{app:models}).

It is worth emphasizing, however, that we cannot rule out a small fraction of the red light coming from a massive galaxy with an older dusty stellar population. This additional stellar component would contribute a lot more mass, potentially amounting to a stellar mass at the order of $10^{10}~\msun$, without violating the main conclusions of a prominent AGN.

The flat (in $F_{\nu}$) spectral shape in the mid-IR also presents a significant challenge to our composite AGN and host galaxy picture since a near-ubiquitous component of AGN at all redshifts is a hot dusty torus that begins to dominate AGN emission longward of 2~$\micron$. That component is not observed in this source. In fact, a much larger sample of sources detected in the JADES fields tells the same basic story; while the rest-frame optical appears to require significant obscuration, there is no evidence of emission from the torus itself out to $2-5~\micron$ in the rest frame \citep{Williams2023,Perez-Gonzalez2024}. %This argues against geometrical effects contributing to a lack of hot dust emission. 
In the majority of cases, with the exception of one object from \citet{Matthee2023:lrd}, this larger sample do not have measurements of broad H$\alpha$, and so \rd\ provides a special case where we can strongly argue for an AGN component despite the lack of hot dust, perhaps refuting the assumption made in \citet{Perez-Gonzalez2024} that these sources are best explained by dusty star formation in the majority of cases. However, \rd\ provides clear confirmation that it is possible to both have an unambiguous broad-line AGN dominate the red continuum in the rest frame optical and simultaneously lack the rising hot dust continuum in the mid-IR.

There is another major puzzle in the SEDs of the LRDs, which is their apparent lack of X-ray emission. Roughly speaking, \rd\ is at least 100 times fainter in the X-ray than we would naively expect from the dereddened $L_{5100}$. The X-ray weakness may be an indication of a high Eddington ratio (e.g., \citealt{Stern2015,Chen2017,Toba2019,Yamada2021}). Interestingly, the [O~\textsc{i}] detections in the Prism spectrum of \rd\ are consistent with the predication from modeling super-Eddington accreting black holes surrounded by dense gas at $z \gtrsim 8$ \citep{Inayoshi2022}.

Put together, the dusty optical continuum, lack of hot dust, and lack of X-ray emission mean that while we can fit the observed UV-optical SED, we do not yet have a full understanding of the underlying physical picture. Thus, measurements of the extinction, intrinsic AGN luminosity, and black hole mass should all be viewed as contingent until a more complete picture can be built.

Even considering a wider redshift range, there are limited number of known analog AGNs with this particular set of spectral characteristics. A handful of cases with a deficiency of hot dust emission are known among high-redshift QSOs \citep{Jiang2010}. These are thought to have evolved relatively dust-poor due to their early formation, although similar hot dust deficient AGN also exist at low-z \citep{Hao2011,Lyu2017,Brown2019,Son2023}. No consensus yet exists about the physical origin of the hot dust deficiency, although a potential explanation could be a difference in torus structure \citep[see discussion in][]{LyuRieke2022}. However, known dust-deficient AGNs are also typically blue in the rest-frame optical reflecting low dust obscuration, in stark contrast to the extreme red rest-optical colors of our confirmed \rd\ and LRDs in the literature. Meanwhile, analogous low-luminosity AGN with similarly v-shaped SEDs in the rest-optical/UV (e.g., blue dust obscured galaxies; \citealt{Noboriguchi2023}) tend to get hot quickly, and are heavily obscured at rest-frame $<6~\mu$m, leading to steeply rising SEDs into the mid-IR.
While it could be possible to reconcile these inconsistent SED properties for specific host geometries that strongly attenuate the rest-optical without an obscuring dust torus (e.g., through polar dust instead; \citealt{Asmus2016,Stalevski2019}), the fact that statistical samples of LRDs in the literature also exhibit flat mid-IR SEDs \citep{Williams2023} argues against orientation or geometrical effects giving rise to such unusual SEDs. Thus, to our knowledge, no known class of AGN across redshifts exhibit similar properties to \rd\ and other LRDs.

\subsection{Black-hole--galaxy Scaling Relations}

Other LRD observations have raised an additional challenge in terms of black hole to galaxy scaling relations. Because the sources are selected to be compact, even at high density there is a limit to their stellar mass. Providing our dust corrections are correct, we infer high ratios of black hole to galaxy mass. However, because of the high spectral resolution observations of \rd, we have an estimate of $\sigma_*$ based on the narrow line widths. Much like the broad-line sources highlighted in \citet{Maiolino2023}, we see some evidence that the black hole mass is in accord with local $M_{\rm BH}-\sigma_*$ scaling relations, even if the stellar mass has not yet caught up. With only a handful of objects it is hard to draw any concrete conclusions yet, but perhaps these relations support the suggestion of \citet{Silk2024} that the high densities in these early galaxies trigger black hole growth and star formation.

\subsection{An Ionized Outflow Traced by He~I Absorption}

We have reported a detection of an absorption feature in the wings of \he\,$\lambda$\,1.083\,$\mu$m in the G395M spectrum. Being in the rest-frame near-IR, studies on this line are scarce, even in nearby objects \citep{Hutchings2002,Leighly2011,Zhang2017,Landt2019,Wildy2021}. At higher redshifts, outflows in AGN have mainly been probed in the highly ionized phase (e.g., C\,{\textsc iv} absorption or \oiii\ emission), while results on neutral gas emission (\cii) have yielded ambiguous results \citep[e.g.,][]{Bischetti2019,Novak2020ApJ}. Our observation thus offers a new look at the outflow properties of LRDs.

The ionized gas outflow in \rd\ has a similar velocity to emission line outflows detected in AGN host galaxies at $z\sim$~2 \citep[e.g.,][]{ForsterSchreiber2019}, but the physical scale of the outflow appears to be much smaller ($\lesssim$~100 pc). We note that the size of the outflow is poorly constrained in our analysis because the \he\ absorption strength is virtually independent of the gas density \citep[e.g.,][]{Ji2015}.

The most powerful outflow in our model grid has a kinetic luminosity $<$~1\% of the bolometric luminosity ($\sim 10^{46}~ {\rm erg \, s^{-1}}$). This suggests that this outflow alone is unlikely to be a significant source of feedback, including regulating accretion onto the black hole or star formation in the host galaxy, but there may be additional outflow mass hidden in other phases \citep{Belli2023}.

In the future, observations covering the H$\alpha$ or H$\beta$ line at higher spectral resolution would help to better constrain the gas density in the absorbing material \citep[e.g.,][]{Wildy2016}. Measurements of absorption column densities in lower ionization absorption lines such as Na~\textsc{i d} would also provide stronger constraints on the ionization structure within the absorbing medium.
In addition, higher resolution observations covering \oiii\ $\lambda$ 5007 \citep{Liu2013} would allow us to search for larger scale ionized outflows manifesting as broad forbidden emission line components and examine the relationship between these kpc-scale outflows and the smaller \he\ outflow seen in our observations.

Intriguingly, a similar absorption feature is found in the G395M spectrum of  GN-28074 \citep{Juodzbalis2024}, an AGN at $z = 2.26$ from the JADES survey \citep{Eisenstein2023}. The observed SED shares similar features as \rd, although the UV coverage is lacking due to its lower redshift. Unfortunately, \he\ becomes unobservable in JWST/NIRSpec at $z>3.5$. However, if a systematic search at lower redshift later reveal that \he\ absorption is a common feature among LRDs, the implication that LRDs are driving potentially vast outflows of ionized gas would be an important one.

\subsection{Final Remarks}

The red objects discovered by JWST continue to challenge the standard views surrounding the evolution of black holes and galaxies. This paper presents an in-depth analysis of an LRD that has both broad lines and JWST/MIRI detections, an occurrence that was observed only once previously. 
The lower redshift of \rd\ provides a better coverage at the rest-frame mid-IR, making the lack of torus emission more evident.
While it remains uncertain whether \rd\ represents the typical properties of LRDs in terms of the inferred characteristics of the host galaxy, AGN, and outflows, this work demonstrates a promising prospect for understanding the nature of the red sources. The RUBIES program will obtain a statistical sample in the near future, and we plan to extend our modeling framework to perform a systematic investigation, which will contribute to an important chapter of the narrative of black hole growth.

\section*{Acknowledgments}
B.W. and J.L. acknowledge support from JWST-GO-04233.009-A.
R.L.D. is supported by the Australian Research Council through the Discovery Early Career Researcher Award (DECRA) Fellowship DE240100136 funded by the Australian Government.
T.B.M. was supported by a CIERA postdoctoral fellowship.
The Cosmic Dawn Center is funded by the Danish National Research Foundation
(DNRF) under grant \#140.
This research was supported by the International Space Science Institute (ISSI) in Bern, through ISSI International Team project \#562 (First Light at Cosmic Dawn: Exploiting the James Webb Space Telescope Revolution).
The JWST data presented in this article were obtained from the Mikulski Archive for Space Telescopes (MAST) at the Space Telescope Science Institute. The specific observations analyzed here can be accessed via\dataset[DOI: 10.17909/c3t4-9p39]{gttps://doi.org/10.17909/c3t4-9p39}.
Computations for this research were performed on the Pennsylvania State University's Institute for Computational and Data Sciences' Roar supercomputer. This publication made use of the NASA Astrophysical Data System for bibliographic information.

\facilities{HST (ACS, WFC3), JWST (NIRCam, NIRSpec), Spitzer (MIPS), Herschel (PACS, SPIRE)}
\software{
Astropy \citep{astropy2013,astropy2018,2022ApJ...935..167A},  
Cloudy \citep{Ferland:2017}, 
emcee \citep{emcee},
dynesty \citep{Speagle2020}, 
Matplotlib \citep{2007CSE.....9...90H}, 
msaexp \citep{Brammer2022}, 
NumPy \citep{2020Natur.585..357H}, 
NUTS \citep{hoffman2014,phan2019},
photutils \citep{Bradley2023},
Prospector \citep{Johnson2021},
pysersic \citep{Pasha2023}, 
SciPy \citep{2020NatMe..17..261V}
}

\section*{Appendix}
% \counterwithin{figure}{section}
% \counterwithin{table}{section}
\renewcommand{\thesection}{\Alph{section}}
\setcounter{section}{0}

\section{Morphology\label{app:morph}}

\begin{figure*} 
\gridline{
  \fig{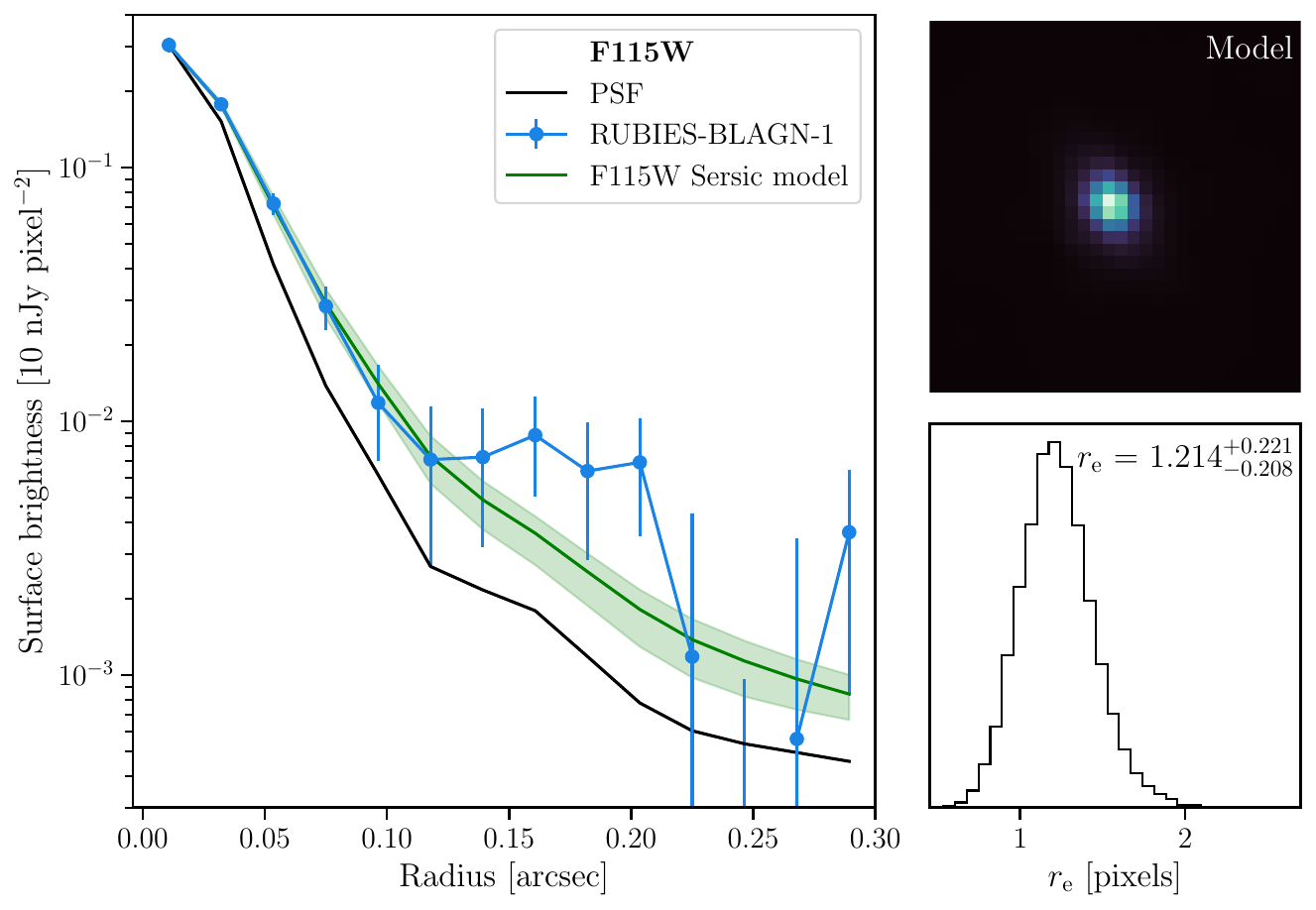}{0.45\textwidth}{(a)}
  \fig{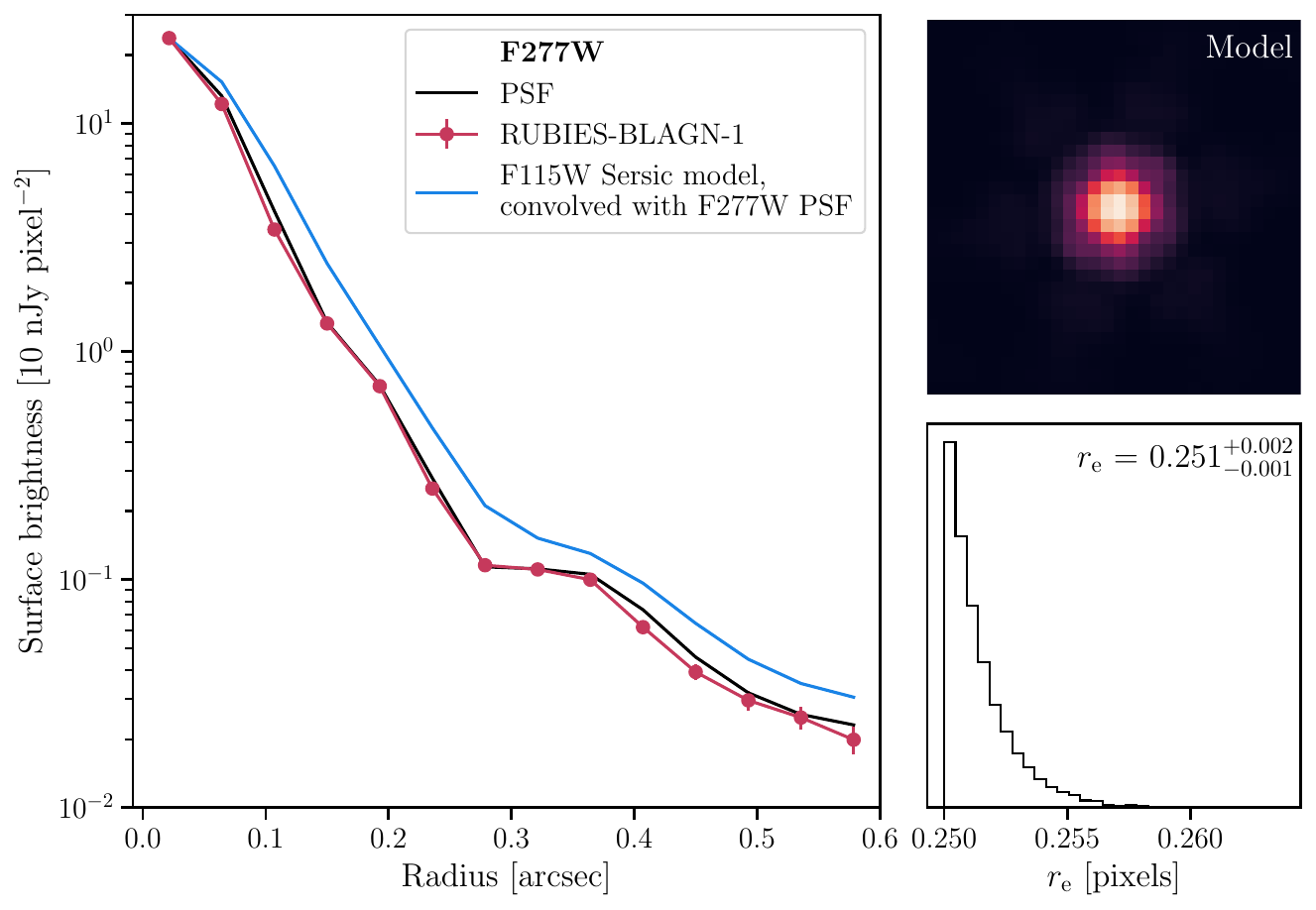}{0.45\textwidth}{(b)}
} 
\caption{Morphological modeling. (a) The observed and modeled radial surface brightness profiles of \rd\ in F115W are shown in blue and green, respectively. The PSF profile, scaled to the source, is over-plotted in black. The upper right panel shows the S\'ersic model fit. The lower right panel shows the marginalized posterior distribution of the effective radius in pixels; $r_{\rm e}=1.21^{+0.22}_{-0.21}$ pixel corresponds to $0.19^{+0.03}_{-0.03}$ kpc.
(b) Same as the left panel, but for F277W. $r_{\rm e}=0.25$ pixel here corresponds to $0.08$ kpc. In blue, we also show the surface brightness profile measured by sampling from the posteriors of the S\'ersic fit in F115W and convolving with the F277W PSF. The median is plotted as a blue curve, with shaded regions indicating the $1\sigma$ uncertainty, and demonstrates that \rd\ is slightly extended in the rest-frame UV, but unresolved at longer wavelengths where emission from the AGN dominates.
}
\label{fig:morph}
\end{figure*}

We perform a morphological analysis of \rd\ in the short wavelength filter F115W and the long wavelength filter F277W, which probe the rest-frame UV ($\sim 0.28~\mu$m) and rest-frame optical ($\sim 0.68~\mu$m), respectively. 
We create a custom mosaic for the F115W imaging, largely following the procedure described in Section~\ref{sec:data}, but drizzled to an improved pixel scale of 0.02\arcsec. 
Empirical PSFs are constructed from non-saturated stars on the 0.02\arcsec\ mosaic for F115W, and the 0.04\arcsec\ mosaic for F277W \citep{Weibel2024}.

We first measure the radial surface brightness profiles using the \texttt{photutils.profiles} module \citep{Bradley2023} with circular apertures that are one pixel wide, ranging from $0-15$ pixels. As seen from Figure~\ref{fig:morph}, the surface brightness profile of \rd\ extends beyond the profile of the PSF in the rest-frame UV (F115W), indicating that it is indeed spatially extended.
Conversely, in F277W the brightness profile of \rd\ and the PSF show no significant differences, indicating that it is a point source at longer wavelengths.

Second, we utilize \texttt{pysersic} \citep{Pasha2023} to fit a S\'ersic profile \citep{Sersic1968} to the images. We assume a uniform prior for the effective radius (between 0.25 to 10 pixels) and a uniform prior on S\'ersic index (between 0.65 to 4). We use the No-U-Turn Sampler (NUTS; \citealt{hoffman2014,phan2019}) to explore the posterior distribution of the morphological parameters with four chains and one thousand warm-up steps, and then five thousand sampling steps for each filter.

Consistent with our findings based on the surface brightness profiles, we find \rd\ to be marginally resolved with an effective radius $r_{\rm e} = 1.21 \pm 0.21\ {\rm pixel}\ (0.024 \pm 0.004\arcsec)$ in the F115W filter. The S\'ersic index is largely unconstrained but the posterior displays a slight preference for $n > 2$. Focusing on the $30 \times 30$ pixel region centered on \rd, the maximum a-posteriori S\'ersic model shows a lower $\chi^2$ by 15\% compared to the value when modeling it solely as a point source. In F277W, \rd\ is unresolved with an inferred $r_{\rm e} \le 0.3 \ {\rm pixel}\ (0.012\arcsec)$. The S\'ersic model shows no improvement in the $\chi^2$ values compared to modelling it as a point source. Both S\'ersic models and marginalized posteriors for the effective radii are included in Figure~\ref{fig:morph}.

Furthermore, we draw 500 posterior samples from the F115W S\'ersic fit, convolve the model images with the F277W PSF, and measure the surface brightness profiles. The median and the $1\sigma$ widths are over-plotted in blue in Figure~\ref{fig:morph}\,(b). The resulting model profile is slightly extended compared to the PSF: therefore, if the rest-frame optical light of \rd\ were as extended as the rest-frame UV, then it should also be marginally resolved in F277W. Crucially, we do not observe such a difference, which implies that the UV and rest-frame optical indeed trace different physical components.

To summarize, we find that \rd\ is marginally resolved in F115W, but unresolved in F277W. 
This analysis is independent of our SED modeling, but, interestingly, consistent with our SED modeling results (\S\ref{subsec:res:fid}): the emission in the rest-frame UV originates from the host galaxy and thus is extended spatially, whereas the rest-frame optical emission is dominated by the AGN that is point-like in nature.

\section{Mock Tests\label{app:mock}}

We generate mock photometric and JWST/Prism data using \prospector\ by drawing random realizations from the prior distributions. We then fit the simulated dataset with the same fiducial model. While this represents an ideal scenario without any model mismatch, the results are instructive in demonstrating the accuracy of the AGN and stellar light decomposition.

\begin{figure*} 
\gridline{
  \fig{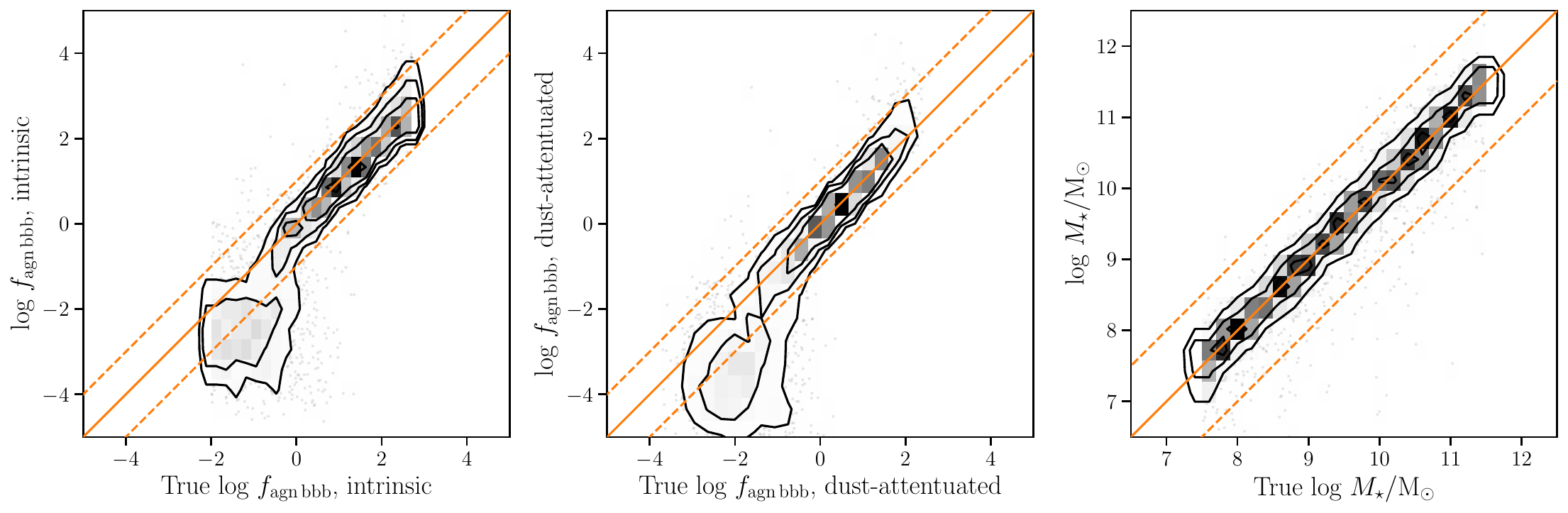}{0.95\textwidth}{(a)}
}
\gridline{
  \fig{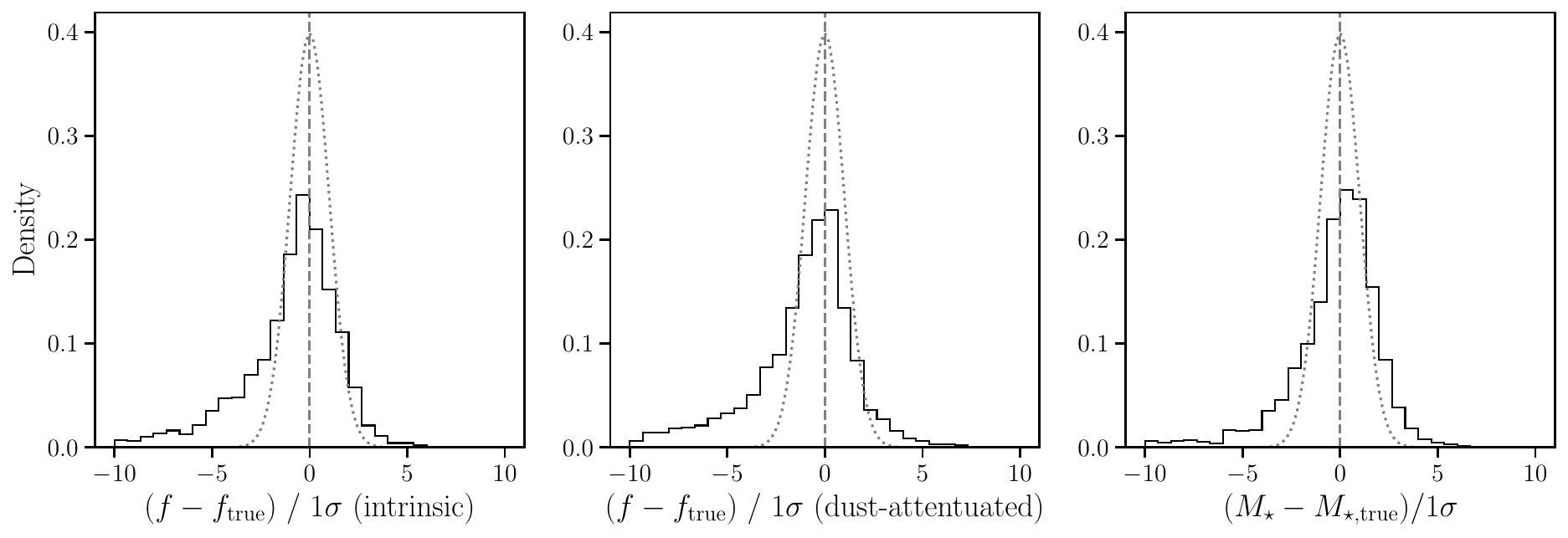}{0.95\textwidth}{(b)}
} 
\caption{Mock tests demonstrating the recovery for the key parameters of interest in our AGN/galaxy composite model.
(a) The inferred values of the fractional continuum contribution of the emission from an AGN accretion disk, pre-dust-attenuation, are plotted against the true values. The solid orange line represents perfect recovery, while the dashed orange lines indicate $\pm 1$ dex to guide the eye. The other two panels show the inferred values of dust-attenuated fractional AGN contribution and galaxy mass, respectively.
(b) Each histogram shows the distribution of residuals normalized by $1\sigma$ uncertainties, in the same order. Unit Gaussian distributions are over-plotted as dotted curves. 
Notably, the fractional AGN contributions for the bright AGN cases are well-recovered within 1 dex, and all the galaxy masses are well-recovered within 0.5 dex.
}
\label{fig:app:mock}
\end{figure*}

As shown in Figure~\ref{fig:app:mock}, both the intrinsic and dust-attenuated fractional AGN continuum contributions, $f_{\rm agn , bbb}$, are recovered within a scatter of $\sim 1$ dex, and the uncertainties are in general well-calibrated, with residuals falling within $3 \sigma$. 
The galaxy masses are recovered with a scatter of $\sim 0.5$ dex, and the residuals are mostly within the $3 \sigma$ uncertainty range.

The clustering at log~$f_{\rm agn , bbb}<0$ in the 2D histograms suggests that the model may underestimate the AGN contribution when dealing with a dim AGN. Notably, this is not the case for \rd, where the broad emission lines indicate a bright AGN, with an inferred $f_{\rm agn , bbb} > 10$. As a result, the mock test provides some confidence in the fiducial understanding of \rd\ presented in this paper.
It is important to note, however, that this confidence statement is only valid if the key modeling assumptions align with the true physical nature of LRDs. As discussed throughout the paper, there is mounting evidence that LRDs are not typical AGNs.

\section{Alternative SED Models\label{app:models}}

Four alternative SED models are considered to offer insight into the systematic uncertainties. The inferred key parameters from each model are summarized in Table~\ref{tab:app:theta}. The galaxy-only model, which is used to argue in favor of our fiducial composite model in Section~\ref{subsec:res:galonly}, is not discussed again here.

\begin{deluxetable*}{lllll}
\tablecaption{Inferred Key Parameters Assuming Alternative SED Models\label{tab:app:theta}}
\tablehead{
\colhead{Parameter} & \colhead{Galaxy-only} & \colhead{AGN-only} & \colhead{Two-component Dust} & \colhead{UV Continuum Masked}
}
\startdata
log~$M_{*}/\msun$ & $11.05_{-0.08} ^{+0.08}$ & -- & $9.53_{-0.18}^{+0.22}$ & $9.50^{+0.20}_{-0.24}$ \\
Age [Gyr] & $0.80_{-0.19}^{+0.11}$ & -- & $0.74^{+0.10}_{-0.11}$ & $0.68^{+0.19}_{-0.19}$ \\
SFR [$\msun\,{\rm yr}^{-1}$] & $164.20 _{-24.39}^{ +41.77} $ & -- & $1.47_{ - 0.60} ^{+ 1.60}$ & $2.76^{+1.98}_{-1.17}$ \\
log~sSFR/${\rm yr}^{-1}$ & $-8.84_{- 0.10}^{ + 0.13}$ & -- & $-9.32_{- 0.23}^{+ 0.18}$ & $-9.00^{+0.24}_{-0.33}$\\
log~$Z/{\rm Z}_\odot$ & $-1.27_{-0.14}^{+0.17}$ & -- & $-1.65_{-0.24}^{+0.36}$ & $-0.93^{+0.55}_{-0.48}$ \\
$\hat\tau_{\rm dust,1}$ & $1.23_{-0.33}^{+ 0.37}$ & -- & $2.50_{-1.36}^{ + 0.96}$ & -- \\
$f_{\rm obrun}$ & $0.47^{+0.31}_{-0.18}$ & -- & $0.32_{- 0.22}^{+0.33}$ & -- \\
$\hat\tau_{\rm dust,2}$ & $2.64_{-0.16}^{+ 0.18}$ & $3.14_{-0.24}^{+0.17}$ & $2.62_{-0.06}^{+0.06}$ & $0.90^{+0.48}_{-0.31}$ \\
$n_{\rm dust,2}$ & $-0.34_{-0.14}^{+ 0.16}$ & $0.02_{-0.02}^{+ 0.03}$ & $-0.97_{ - 0.02}^{+ 0.03}$ & $-0.86^{+0.20}_{-0.10}$ \\
$f_{\rm no dust}$ & $0.01_{- 0.00}^{+0.00}$ & $0.50_{- 0.26}^{+ 0.19}$ & $0.59_{- 0.26}^{+ 0.24}$ & $0.37^{+0.34}_{-0.25}$ \\
$\log f_{\rm AGN, 7500A}$ & -- & -- & $1.13_{-0.16}^{+ 0.10}$ & $0.97^{+0.16}_{-0.15}$ \\
$\log f_{\rm AGN, torus}$ & -- & -- & $-2.50_{-0.13}^{ + 0.63}$ & $-2.31^{+0.22}_{-0.02}$ \\
$\hat\tau_{\rm torus}$ & -- & -- & $7.66_{- 2.06}^{ + 36.20}$  & $9.86_{-3.57} ^{+9.14}$ \\
$\hat\tau_{\rm dust,4}$ & -- & $0.52_{-0.18}^{ + 0.29}$ & -- & $2.20^{+0.39}_{-0.47}$ \\
$n_{\rm dust,4}$ & -- & $-1.07_{-0.13}^{+ 0.12}$ & -- & $-1.43^{+0.26}_{-0.20}$ \\
\enddata
\end{deluxetable*}

\subsection{AGN Only}

\begin{figure*} 
\gridline{
  \fig{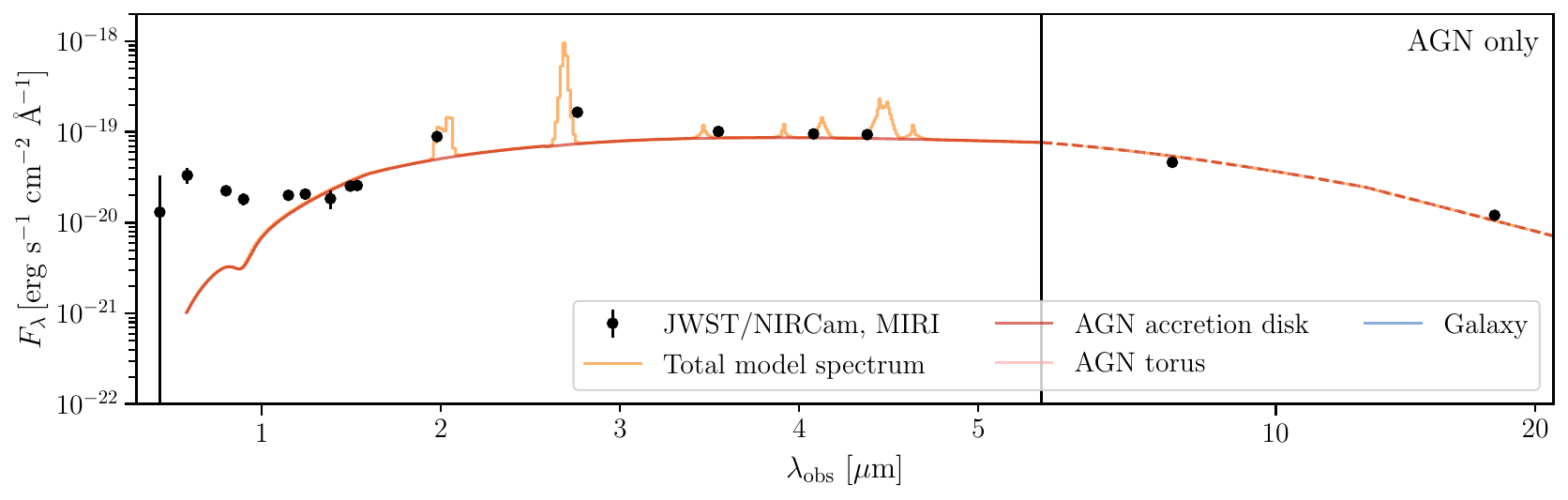}{0.7\textwidth}{(a)}
} 
\gridline{
  \fig{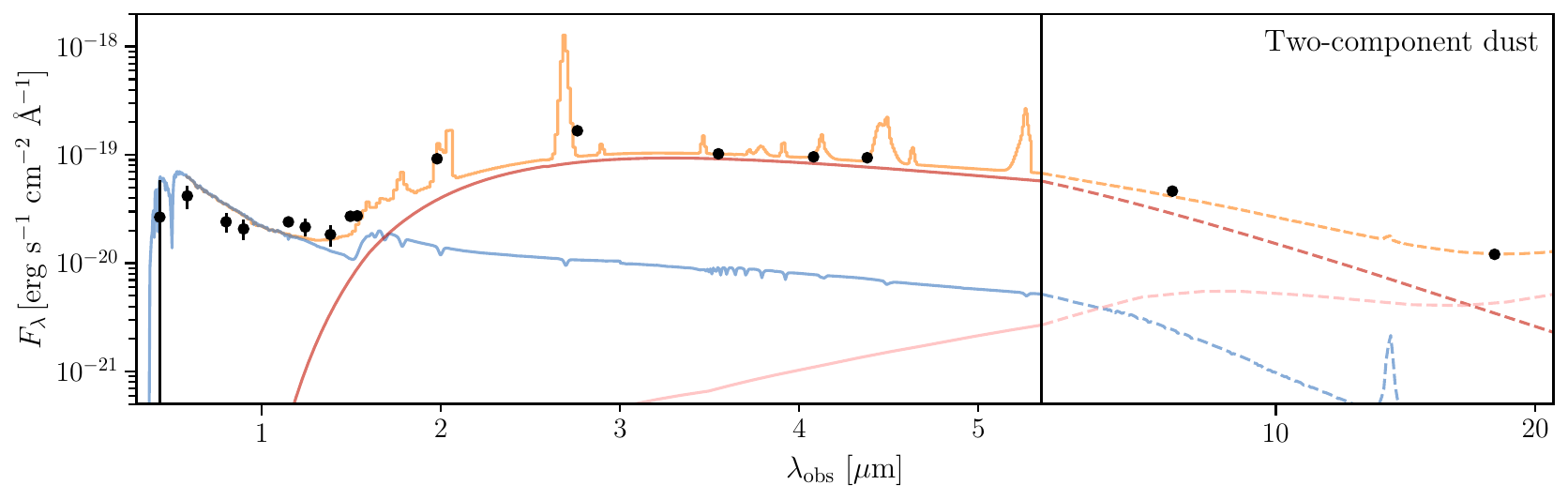}{0.7\textwidth}{(b)}
} 
\gridline{
  \fig{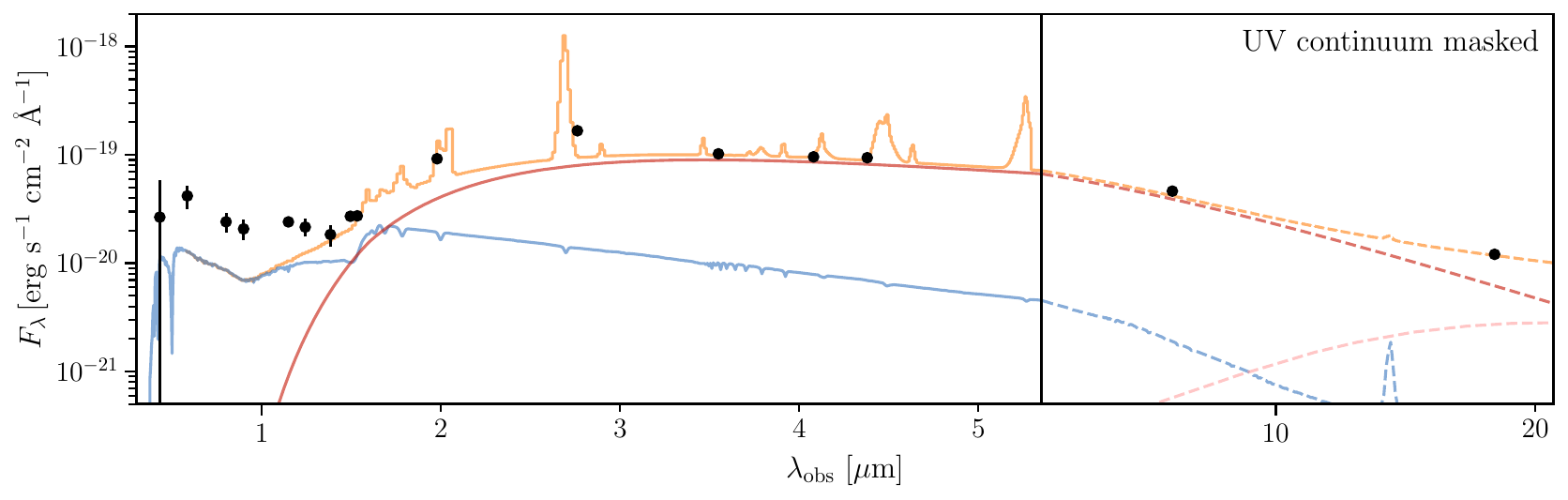}{0.7\textwidth}{(c)}
} 
\caption{Alternative SED models. (a) Fitting with AGN light only.
(b) Fitting with the standard two-component dust model.
(c) Fitting only the rest optical spectrum using the fiducial model.
In each panel, the best-fit total model spectrum, with the marginalized emission lines added, is plotted in orange. The model spectra of the AGN accretion disk, torus, and the galaxy are plotted in red, pink, and blue, respectively. The photometric observations are included as black dots, but the observed spectrum is omitted for clarity.
}
\label{fig:app:models_other}
\end{figure*}

We construct a simple AGN model, composed solely of accretion disk and torus emissions.
The best-fit is shown in Figures~\ref{fig:app:models_other}\,(a). 
This AGN-only model fails to describe the blue UV slope, as expected given the intrinsic spectral shape of the AGN accretion disk. Combining this result with the galaxy-only model supports the inference from our fiducial model that the observed spectrum of \rd\ comprises light from the AGN and its host galaxy. We note, however, that a pure-AGN model with a scattered light component \citep{Greene2023} remains a possibility.

\subsection{Dust}

The redness of \rd\ does not necessarily mean that the AGN receives additional attenuation. We thus consider the standard dust model, where dust reddening of the stellar light is described by two components \citep{Charlot2000} with a flexible dust attenuation curve \citep{Noll2009}.
As in the fiducial model, a fraction of the stellar light is allowed to be seen outside the diffuse dust. In addition, a fraction of the young stellar light is allowed to be not attenuated by the birth cloud, which represents runaway OB stars or escaping ionizing radiation.

This model infers significant dust presence, driven by the red continuum as in the fiducial model. However, without the additional attenuation experienced only by the AGN, $\sim 30$\% of the young stars and $\sim 60$\% of all stars are required to be located outside the birth cloud and the diffuse dust, respectively, to produce the UV excess. It is less intuitive to reconcile this situation where a significant fraction of the stars remains unaffected by dust attenuation while the AGN experiences substantial attenuation.

\subsection{Fitting the Rest Optical Spectrum Only}

It is possible that the UV continuum is of non-stellar origin. We test this hypothesis by masking the UV, and fit the red continuum with the fiducial model. This setting fails to predict an UV excess, meaning that the UV continuum does not have to be composed of starlight. It can, for example, be the scattered light from the AGN \citep{Greene2023}. Crucially, the inferred stellar mass in this case is about the same as that from the fiducial model, suggesting that the UV continuum has less constraining power. While we can add in a non-stellar component for the UV, this result means that the extra UV component will not add much information under the assumed dust model, except improving the $\chi^2$.

\bibliography{rubies_brd_wang.bib}

\end{document}